\documentclass{amsart}
\usepackage{amssymb}
\usepackage{amsmath}
\usepackage[inline]{asymptote}

\newtheorem{lemma}{Lemma}
\newtheorem{conjecture}{Conjecture}
\newtheorem{proposition}{Proposition}
\newtheorem{observation}{Observation}
\newtheorem{corollary}{Corollary}
\newtheorem{definition}{Definition}
\newtheorem{example}{Example}

\newenvironment{Proof}
{\begin{trivlist}\item[]{{\sc Proof.}}}{\hfill{$\square$}\noindent\end{trivlist}}

\def\voter{player}
\def\BZ{\operatorname{Bz}^r}
\def\PGI{\operatorname{PGI}^r}
\def\Shift{\operatorname{S}^r}
\def\J{\operatorname{Jo}^r}
\def\DP{\operatorname{DP}^r}
\def\SDP{\operatorname{SDP}^r}

\usepackage{color}

\begin{document}

\title{The cost of getting local monotonicity}
\author{Josep Freixas}
\address{Departament de Matem\`{a}tica Aplicada 3 i Escola Polit\`{e}cnica Superior d'Enginyeria de Manresa 
(Universitat Polit\`{e}cnica de Catalunya). Spain. josep.freixas@upc.edu \,\,\, \thanks{Research partially supported by ``Ministerio de Econom\'{\i}a y Competitividad proyecto MTM2012-34426/FEDER".}}
\author{Sascha Kurz}
\address{Department of Mathematics, University of Bayreuth, 95440 Bayreuth, Germany.\\ Tel.: +49-921-557353, Fax: +49-921-557352, sascha.kurz@uni-bayreuth.de }
\begin{abstract}
  In \cite{holler1982forming} Manfred Holler introduced the Public Good index as a proposal to divide a public good among players. In
  its unnormalized version, i.e., the raw measure, it counts the number of times that a player belongs to a minimal winning coalition.
  Unlike the Banzhaf index, it does not count the remaining winning coalitions in which the player is crucial. Holler noticed
  that his index does not satisfy local monotonicity, a fact that can be seen either as a major drawback \cite[221 ff.]{0954.91019}
  or as an advantage \cite{holler2004monotonicity}.

  In this paper we consider a convex combination of the two indices and require the validity of local monotonicity. We prove that the
  cost of obtaining it is high, i.e., the achievable new indices satisfying local monotonicity are closer to the Banzhaf index than to
  the Public Good index. All these achievable new indices are more solidary than the Banzhaf index, which makes them as very suitable
  candidates to divide a public good.

  As a generalization we consider convex combinations of either: the Shift index, the Public Good index, and the Banzhaf index, or alternatively:
  the Shift Deegan-Packel, Deegan-Packel, and Johnston indices.
  
  \medskip
  
\end{abstract}

\maketitle

\section{Introduction}

\noindent
Consider a set of players who jointly make decisions under a given set of rules. Here we  specialize to simple games and
subclasses thereof. Power indices address the question of how much power collective decision rules, like  a weighted
(voting) rule, award to each individual player: is player $i$ more or less powerful than player $j$, and by how much?
For an example of an applied voting power analysis in the EU, we refer the interested reader 
to e.g.\ \cite{EU_power_distribution,EU_enlargement,mika1994EC}.

Different power indices measure different aspects of power and there is still a lot of research in order to answer the
question which index to choose, see e.g.\ \cite{holler2013reflections}. For a recent overview of different power indices
see e.g.\ \cite{bertini2013comparing}. Many of these indices are based on decisiveness.  A player is called decisive in a
coalition if his/her deletion in the coalition changes its status from winning to losing, so that the individual is
decisive or crucial for it. All power indices, the classical and the newly introduced ones, considered in this paper are
indeed based on counting different types of decisiveness for players in coalitions.

Some particular rules, weighted games, specify that each player $i=1,\dots,n$ has a specific voting weight $w_i$ and that
a collective decision requires enough supporters such that their total weight equals or surpasses a decision quota $q$.
Let $p_i$ be the power value assigned to player $i$ by a power index. The power index is called \emph{locally monotonic} if
$w_i\ge w_j$ implies $p_i \geq p_j$, i.e., a {\voter}~$i$ who controls a large share of vote does not have less power than a
{\voter}~$j$ with smaller voting weight. Local monotonicity is considered as an essential requirement for power measures by many authors.
Felsenthal and Machover~\cite[221 ff.]{0954.91019}, for instance, argue that any a priori measure of power that violates local
monotonicity, LM for brevity, is `pathological' and should be disqualified as serving as a valid yardstick for measuring power.
On the other hand e.g.\ in \cite{holler2004monotonicity} it is argued that local non-monotonicity is a very valuable property
of a power index, since it can reveal certain properties of the underlying decision rule that are overlooked otherwise.

Local monotonicity is an implication of the \emph{dominance} postulate which is based on the \emph{desirability} relation
as proposed by Isbell~\cite{0083.14301}. This property formalizes that a {\voter}~$i$ is at least as desirable as a {\voter}~$j$
if for any coalition $S$, such that $j$ is not in $S$ and the union of $S$ and $\{j\}$ is a winning coalition, i.e., is able
to pass the collective decision at hand, the union  of $S$ and $\{i\}$ is also a winning coalition. A power index satisfies
\emph{dominance} if $p_i \geq p_j$ whenever $i$ dominates $j$.

Freixas and Gambarelli~\cite{freixas1997common} use desirability to define reasonable power measures and note that the
dominance postulate implies local monotonicity. In this paper we will consider the Public Good, the Banzhaf, the Shift,
the Shift Deegan-Packel, the Deegan-Packel, the Johnston index and convex combinations thereof. Since the Deegan-Packel
index~\cite{deegan1978new}, and the Public Good Index (see Holler~\cite{holler1982forming}; Holler and Packel~\cite{holler1983})
violate local monotonicity, they also violate the
dominance postulate. Moreover, any violation of local monotonicity for the Deegan-Packel index implies a violation of the
Shift Deegan-Packel index (see~\cite{AlonsoMeijide20123395}) and any violation of the local monotonicity for the Public Good
Index implies a violation of the Shift index (see~\cite{alonso2010new}). It is well-known that the Banzhaf~\cite{banzhaf1964weighted}
and Johnston~\cite{johnston1978measurement} indices satisfy the dominance postulate and therefore local monotonicity.
If one or several power indices violate LM then a convex combination with another power index, that does not violate
LM, yields a power index that also does not violate LM as long as the weight of the latter index is large enough. To
study how large this has to be is the purpose of this paper.

Some works are devoted to verify the properties of dominance or local monotonicity (among others) for some power indices and
to show failures for some other power indices (see among others, Felsenthal and Machover~\cite{felsenthal1995postulates} or
Freixas et al.~\cite{freixas2012ordinal}). Other works are devoted to study subclasses of games for which a given power index
not fulfilling local monotonicity satisfies it for such a subclass of games (see for instance, Holler et
al.~\cite{holler2001constrained} and Holler and Napel~\cite{holler2004monotonicity} for the Public Good Index). 
Here we will also make a new contribution of this type, i.e., we consider two new subclasses of games for which the Public Good Index 
satisfies local monotonicity.

The strictest generalization of local monotonicity is proportionality of power and weights. For the classical
power indices this property is satisfied for a subset of weighted games only. Power indices which
generally satisfy this property are constructed in \cite{average_representation}.

This paper starts by modifying the Public Good index with the purpose to achieve a new power index being local monotonic and
more solidary than the Banzhaf index. These two properties make those achievable power indices (if they exist) well-situated
as yardstick for doing a fair division of a public good. The idea of such modification is nothing else than an hybrid between
the original Public Good index and the Banzhaf index. It will turn out that the cost of obtaining local monotonicity is
rather high, i.e., the achievable new indices satisfying local monotonicity are closer to the Banzhaf index than to the Public
Good index. However these indices stress more in minimal winning coalitions, as the Public Good index does, than in the rest of crucial
winning coalitions, with goes in the direction of Riker's size principle, see \cite{riker1962theory}. The final result permits
to find new indices being locally monotonic and being more solidary than the Banzhaf index, which makes them as good alternatives
for the fair division of a public good among participants in the voting procedure.


The idea developed previously naturally extends when the raw Shift index is incorporated to the duo formed by the raw Public Good and
raw Banzhaf indices. Local monotonic indices which are convex combinations of the three given raw indices are a further target of our research.

As an extension we do a similar study for convex combinations of the raw Johnston index, the raw Deegan-Packel index, and the
raw Shift Deegan-Packel index.

The remaining part of the paper is organized as follows: In Section~\ref{sec_notation} we introduce the basic notation of games and
power indices. Two subclasses of weighted games satisfying local monotonicity are presented in Section~\ref{some_subclasses}. The 
concept of considering convex combinations of some power indices as a new power index is outlined in
Section~\ref{sec_convex_combinations}. The cost of local monotonicity is introduced in the same section. Additionally we prove
some structural results. An integer linear programming approach to compute the cost of local monotonicity is presented
in Section~\ref{sec_ILP}. With the aid of the underlying algorithm we are able to state some exact values and lower bounds
for the cost of local monotonicity in Section~\ref{sec_exact_values_and_bounds}. The set of all convex multipliers leading to
a locally monotonic power index is the topic of Section~\ref{sec_polyhedron}. We end with a conclusion
in Section~\ref{sec_conclusion}.

\section{Notation, games and indices}
\label{sec_notation}

\noindent
In the following we will denote the set of players, which jointly make a decision, by $N$ and assume w.l.o.g.\
that the players are numbered from $1$ to $n$, i.e., $N=\{1,\dots,n\}$. Here we restrict ourselves to binary decisions,
i.e., each player can either vote $1$, meaning `yes', or $0$, meaning `no', on a certain issue. We call a subset
$S\subseteq N$, collecting the `yes'-voters, \emph{coalition}. A (binary) decision rule is formalized as a mapping
$v:2^N\rightarrow\{0,1\}$ from the set of possible coalitions to the set of possible aggregated decisions. It is quite
natural to require that the aggregated decision transfers the players decision if they all coincide and that an enlarged
set of supporters should not turn the decision from yes to no:

\begin{definition}
  A \emph{simple game} is a mapping $v:2^N\rightarrow\{0,1\}$ such that $v(\emptyset)=0$, $v(N)=1$, and $v(S)\le v(T)$ for all
  $S\subseteq T\subseteq N$.
\end{definition}

Having local monotonicity in mind we additionally require that the players are linearly ordered according to
their capabilities to influence the final group decision. This can be formalized, as already indicated in the introduction,
with the desirability relation introduced in \cite{0083.14301}.
\begin{definition}
  We write $i\sqsupset j$ (or $j \sqsubset i$) for two players $i,j\in N$ of a simple game
  $v$ if we have $v\Big(\{i\}\cup S\backslash\{j\}\Big)\ge v(S)$ for all
  $\{j\}\subseteq S\subseteq N\backslash\{i\}$ and we abbreviate $i\sqsupset j$,
  $j\sqsupset i$ by $i\square j$.
\end{definition}
In words we say that $i$ dominates $j$ for $i\sqsupset j$ and we call $i$ and $j$ equivalent iff $i\square j$.

\begin{definition}
  \label{def_complete_simple_game}
  A simple game $v$ is called \emph{complete} if the binary relation $\sqsupset$ is a total
  preorder, i.e.,
  \begin{itemize}
    \item[(1)] $i\sqsupset i$ for all $i\in N$,
    \item[(2)] $i\sqsupset j$ or $j\sqsupset i$ for all $i,j\in N$, and
    \item[(3)] $i\sqsupset j$, $j\sqsupset h$ implies $i\sqsupset h$ for all $i,j,h\in N$.
  \end{itemize}
\end{definition}

We call a coalition $S$ of a simple game $v$ \emph{winning} if $v(S)=1$ and \emph{losing} otherwise. Each simple game
is uniquely characterized by its set $\mathcal{W}$ of winning coalitions (or its set $\mathcal{L}$
of losing coalitions). A winning coalition $S$ such that each of its proper subsets is losing is called a \emph{minimal
winning} coalition. The set $\mathcal{M}$ of minimal winning coalitions is already sufficient to uniquely
characterize a simple game. For complete simple games the defining set of winning coalitions can be
further reduced. A minimal winning coalition $S$ is called \emph{shift-minimal} if for each pair of players $i$, $j$ with
$i\in S$, $j\notin S$, $i\sqsupset j$, $j\not\sqsupset i$ we have $v(S\backslash\{i\}\cup\{j\})=0$, i.e.,
replacing a player by a (properly) dominated player turns the coalition into a losing one. With this, each
complete simple game is uniquely characterized by its set $\mathcal{S}$ of shift-minimal winning coalitions.

A very transparent form of dominance is induced by weights.

\begin{definition}
  A simple game $v$ is called \emph{weighted} (weighted game for brevity) if and only if there exist weights $w_i\in\mathbb{R}_{\ge 0}$,
  for all $i\in N$, and a quota $q\in\mathbb{R}_{>0}$ such that $v(S)=1$ is equivalent
  to $w(S):=\sum_{i\in S} w_i\ge q$ for all $S\subseteq N$.
\end{definition}

Given such a weighted representation we write $v=[q;w_1,\dots,w_n]$. All weighted games are complete. As remarked
before $w_i\ge w_j$ implies that player~$i$ dominates player~$j$, i.e., $i\sqsupset j$, while $i\square j$ is still
possible even for $w_i>w_j$.

In order to measure the influence of the players we use the concept of a \emph{power index}, which we in general consider
as a mapping from a set $\mathfrak{G}$ of games to a vector of $n$ real numbers, where $n$ is the number of players of the
specific game.  
In most applications, considering subsets of the set of simple games, the image is a vector of $n$ non-negative
real numbers upper bounded by $1$.   
To this end we denote by $\mathfrak{S}$ the set of
simple games, by $\mathfrak{C}$ the set of complete simple games, and by $\mathfrak{W}$ the set of weighted games. In order
to stress the underlying class of games, we speak of a power index $P$ \emph{on $\mathfrak{G}$}, whenever it is not clear
from the context.

In some contexts it is appropriate to further restrict the class of games:
\begin{definition}
  \label{def_proper_strong}
  A simple game is called \emph{proper} if the complement $N\backslash S$ of any winning coalition $S$ is losing. It is called
  \emph{strong} if the complement $N\backslash S$ of any losing coalition $S$ is winning. A simple game that is both proper and
  strong is called \emph{constant-sum} (or self-dual, or decisive).
\end{definition}

We will denote the restriction to proper, strong, or constant-sum games by a superscript $\operatorname{p}$, $\operatorname{s}$,
and $\operatorname{c}$, respectively, i.e., we write $\mathfrak{S}^{\operatorname{p}}$, $\mathfrak{S}^{\operatorname{s}}$, and
$\mathfrak{S}^{\operatorname{c}}$ in the case of simple games. If $P$ is a power index on $\mathfrak{G}$, then there is a restricted
power index $P'$ on $\mathfrak{G}'$ for all $\mathfrak{G}'\subseteq \mathfrak{G}$,

Having the general concept of a power index $P$ on $\mathfrak{G}$ at hand, i.e.,  $P(v)=(P_1(v), \dots,$ 
$P_n(v))=(p_1,\dots,p_n) \in \mathbb{R}^n$, we can define the properties that we are interested in this paper:

\begin{definition}
  A power index $P$ on $\mathfrak{G}\subseteq\mathfrak{C}$ satisfies the \emph{dominance property} if we
  have $p_i\ge p_j$ for all complete simple games $v\in\mathfrak{G}$ and all pairs of players $i\sqsupset j$, where
  $P(v)=(p_1,\dots,p_n)$.
\end{definition}

Restricting the dominance property from the class of complete simple games to weighted games, we speak of local monotonicity.

\begin{definition}
  A power index  $P$ on $\mathfrak{G}\subseteq\mathfrak{W}$ satisfies \emph{local monotonicity} (LM) if we have $p_i\ge p_j$ 
  for all weighted games $v\in\mathfrak{G}$ and all pairs of players $i\sqsupset j$ (or $w_i\ge w_j$), 
 where $P(v)=(p_1,\dots,p_n)$.
\end{definition}

We remark that the dominance property for the subclass of weighted games implies local monotonicity and local monotonicity 
implies \emph{symmetry}, i.e., we have $p_i=p_j$ for all $i\square j$. Further properties of
classical power indices where named and studied in the literature, e.g., if all $p_i$'s are non-negative and sum up to one
the power index is called \emph{efficient}. If $p_i=0$ for all players $i$ not contained in any minimal winning coalition, also
called \emph{null players} or \emph{nulls}, then the power index is said to satisfy the \emph{null player property}. Removing
a null player from a simple game (complete simple game or weighted game) $v$ results in a simple game (complete simple game
or weighted game) $v'$ -- more formally $v':=2^{N\backslash\{i\}}\rightarrow\{0,1\}$ $v'(S)=v(S)$ for all
$S\subseteq N\backslash\{i\}$. If $p_j=p'_j$ for all $j\in N\backslash\{i\}$ and $p_i=0$, where
$P(v)=(p_1,\dots,p_n)$ and $P(v')=(p'_1,\dots,p'_{i-1},p'_{i+1},\dots,p'_n)$, we call $P$ \emph{invariant for nulls}.
We call a power index \emph{strictly positive} if $p_i>0$ for all non-null players $i$ and $p_j=0$ for all
null players~$j$.

In order to state the definition for the set of power indices mentioned in the introduction we call a winning coalition $S$
\emph{decisive} for player $i\in S$ if $S\backslash\{i\}$ is losing. Given a simple game $v$, by $\mathcal{D}_i$ we denote
the set of coalitions $\{i\}\subseteq S\subseteq N$ being decisive for player~$i$. Similarly, we denote by $\mathcal{M}_i$
the set of minimal winning coalitions containing player~$i$ and by $\mathcal{S}_i$ the set of shift-minimal
winning coalitions containing player~$i$ (provided that the game is complete). In order to specify a power index $P$
it suffices to define the mapping for each coordinate $P_i(v)=p_i$.

\begin{definition}
  The (raw) Banzhaf index $\BZ$ and the (raw) Public Good index $\PGI$ of a simple game $v$ are given by
  $\BZ_i(v)=\left|\mathcal{D}_i\right|$ and $\PGI_i(v)=\left|\mathcal{M}_i\right|$. The (raw) Shift index of a
  complete simple game $v$ is given by $\Shift_i(v)=\left|\mathcal{S}_i\right|$.
\end{definition}

The (raw) Banzhaf index $\BZ$, the (raw) Public Good index $\PGI$, and the (raw) Shift index $\Shift$ all are based
on decisive coalitions since they count subsets of decisive coalitions containing a given player~$i$. The Banzhaf index
counts all decisive coalitions for each player, while the Public Good index counts only the minimal and the Shift index 
only the shift-minimal ones. Thus, for each player we have the inclusion
$$
  \mathcal{S}_i\subseteq \mathcal{M}_i\subseteq \mathcal{D}_i.
$$
Counting a certain subset of coalitions is the base for many power indices, see e.g.\ \cite{inverse} for further examples.
As defined several coalitions can be counted multiple times, i.e., a minimal winning coalition $S$ is counted for
every player $i$ contained in $S$ in the computation of $\PGI$. If one wants to avoid this characteristic and instead count
each distinguished coalition just once, then one can divide `spoils' of each coalition equally among its decisive
members. This general construction is called \emph{equal division} version of a given power index (based on counting functions)
in \cite{inverse}. Applying this construction on our first set of power indices we obtain the second set:

\begin{definition} \label{d:J-DP-SDP}
  The (raw) Johnston index $\J$ and the (raw) Deegan-Packel index $\DP$ of a simple game $v$ are given by
  $\J_i(v)=\sum_{S\in\mathcal{D}_i} \frac{1}{\# \text{ decisive players in }S}$ and $\DP_i(v)=\sum_{S\in\mathcal{M}_i}\frac{1}{|S|}$.
  The (raw) Shift Deegan-Packel index of a complete simple game $v$ is given by $\SDP_i(v)=\sum_{S\in\mathcal{S}_i}\frac{1}{|S|}$.
\end{definition}

Note that in the two last definitions $|S|=\# \text{ decisive players in }S$, since all players in minimal winning (or in 
shift-minimal winning) coalitions are decisive in them.

\section{Two subclasses where the Public Good and the Deegan-Packel satisfy local monotonicity}
\label{some_subclasses}

As it is well-known the the Banzhaf and the Johnston indices (and the Shapley-Shubik too) satisfy both the dominance property 
and local monotonicity since all of them preserve the desirability relation (see e.g.~\cite{freixas2012ordinal}). The other power 
indices previously introduced in Section~\ref{sec_notation} and extensively analyzed in this paper do not preserve the desirability 
relation and consequently fail to fulfill both the dominance and the local monotonicity properties.

The purpose of this section is to provide subclasses of games, not introduced before, satisfying the dominance property or the 
local-monotonicity for  the Public Good and the Deegan-Packel indices. As seen below the cost for one of these two power indices 
to satisfy the local monotonicity property for a given game is related to the proximity or remoteness of the game to be in these 
subclasses.

\begin{definition}
A simple game is:
\begin{enumerate}
\item \emph{uniform} if all minimal winning coalitions have the same cardinality,
\item \emph{uniformly complete} if it is both complete and uniform,
\item \emph{uniformly weighted} if it is both weighted and uniform.
\end{enumerate}
\end{definition}

As an abbreviation we denote the corresponding subclasses by $\mathfrak{U}$, $\mathfrak{U}_c$, and $\mathfrak{U}_w$, respectively. 
Obviously, a uniformly weighted game is a uniformly complete game and a uniformly complete game is a uniform game, while the 
converses are not true.

The second observation is that if a game is uniform then all shift-minimal winning coalitions have the same cardinality because 
$\mathcal S \subseteq \mathcal M$ for all simple game. However the converse is also true, i.e., if all shift-minimal winning 
coalitions have the same cardinality then all the minimal winning coalitions have the same cardinality too. This is because the 
coalitions in $\mathcal M \setminus \mathcal S$ are obtained from those in $\mathcal S$ by one-to-one replacements of weaker 
players, according to the desirability relation, for stronger players; but these exchanges do not affect the cardinalities of 
the coalitions involved. Hence, we can exchange in previous definition the term ``minimal winning coalition" by ``shift-minimal 
winning coalition".

If $l$ is the cardinality of all minimal winning coalitions in a uniform game, in any of its forms, then 
$\mathcal M = \mathcal M(l)$ where $\mathcal M(l)$ is the set of minimal winning coalitions of cardinality $l$.

%
%

\begin{proposition} \label{P:mainresult}
Let $v$ be a uniform complete simple game, characterized by $\mathcal{W}$ and $N$, then the Public Good and Deegan-Packel 
indices satisfy the dominance property.
\end{proposition}

\begin{Proof}
Since $DP^r_i = PGI^r_i/l$ for all $i \in N$, where $l = \min \{|S| \, : \, S \in \mathcal W \}$, it suffices to prove 
the statement for the Public Good index.


Assume $i \sqsupset j$, then for all $S \subseteq N \setminus \{i,j\}$ with $S\cup\{j\}\in \mathcal{M}$ we 
have $S\cup\{i\}\in\mathcal{W}$. If $S \cup \{i\} \notin \mathcal M$, then there would exist a player $k \in S$ such that $(S
\cup \{i\}) \setminus \{k\} \in \mathcal W$, but $|(S \cup \{i\}) \setminus \{k\}| = |S \cup \{j\}|-1=l-1$, which
is a contradiction with the fact that all minimal winning coalitions have the same size. Thus, $S \cup
\{i\} \in \mathcal M$ and therefore $ \PGI_i=|\mathcal M_i| \geq |\mathcal M_j|=\PGI_j$.
\end{Proof}

We remark that $i \sqsupset j$, $i \not\sqsupset j$ even implies $\PGI_i > \PGI_j$.


\begin{corollary} \label{c:usg} The Public Good and Deegan-Packel indices satisfy dominance and local monotonicity properties on the 
classes $\mathfrak{U}_c$ and $\mathfrak{U}_w$, respectively.
\end{corollary}

As a consequence, for uniform complete simple games the ranking given by the desirability relation 
coincides with the rankings given by Public Good and 
Deegan-Packel indices. So these ranking also coincide with the rankings of the Shapley-Shubik, Banzhaf, and Johnston indices,  
see~\cite{freixas2012ordinal}.


Despite the restrictive definition of uniform complete simple games, their number is large. For instance, there are 
(see e.g.~\cite{1151.91021,kurz2013dedekind}) $2^{n}-1$ uniform complete simple games with just one type of shift-minimal winning 
coalitions of $n$ voters. Table~\ref{table_uniform} provides enumerations for small values of $n$ on the number of
uniform complete games ($\mathfrak{U}_c$) and uniform weighted games ($\mathfrak{U}_w$). For $n=10$ players the number 
of uniform complete simple games is given by $3\,049\,712\,101$ and for $n=11$ players the respective number 
larger than $25\cdot 10^{12}$. Without considering symmetry the 
number of uniform simple games with $n$ players is given by $\sum\limits_{k=1}^{n} 2^{{n\choose k}-1}$.



\begin{table}[htp]
\begin{center}
\begin{tabular}{|c|c|c|c|c|c|c|c|c|c|}
\hline \hline
n & 1 & 2 & 3 & 4 & 5 & 6 & 7 & 8 & 9 \\ \hline \hline
$\mathfrak{U}_c$  & 1 & 3 & 7 & 16 & 41 & 140 & 843 & 14\,434 & 1\,410\,973   \\
$\mathfrak{U}_w$  & 1 & 3 & 7 & 16 & 41 & 125 & 458 & 2\,188 & 20\,079 \\
\hline \hline
\end{tabular}
\caption{Number of uniform complete simple games and uniform weighted games.}
\label{table_uniform}
\end{center}
\end{table}


Being a uniform game is a sufficient condition for both the Public Good and the Deegan-Packel index to be local monotonic. 
However, this condition is not necessary as the following example illustrates.

\begin{example} Let $v$ be the $4$-person game uniquely characterized  by $N = \{1,2,3,4\}$ and
$\mathcal M = \{ \{ 1,2 \}, \{ 1,3 \}, \{ 1,4\}, \{2,3,4\} \}$, i.e., $v$ admits the weighted representation
$[3;2,1,1,1]$. This game is the unique weighted game of $4$ voters which is not uniform but it satisfies the 
dominance property for the Public Good index. 
\end{example}


At the very least, this game is captured by the following two definitions.

\begin{definition}
Let $v$ be a simple game with player set $N=\{1,\dots,n\}$ and
$$ i  \succsim  j \quad \text{if and only if} \quad  \sum_{l=1}^k
|\mathcal{M}_i(l)| \geq  \sum_{l=1}^k  |\mathcal{M}_j(l)|
\quad \text{for all} \quad k=1,2,\dots,n.
$$
\end{definition}
Then $\succsim$ is a preordering on $N$, i.e., a reflexive and transitive relation, called the
\emph{layer relation}.


\begin{definition}
A simple game $v$ on $N$ is \emph{flat} whenever $\succsim$, the layer
relation on $N$, is total.
\end{definition}

Thus, in a flat simple game, we either  have $i \succsim  j$ or $j \succsim i$ for all $i,j \in N$. 
Let $\mathfrak{F}$ be the class of flat games. 

\begin{proposition} 
Let $v$ be a flat complete simple game then the Public Good and Deegan-Packel 
indices satisfy the dominance property.
\end{proposition}

\begin{Proof}
Assume $i \sqsupset j$ and let $k\le n-2$ be the maximal integer such that
$$
  \left\{S\subseteq N\backslash\{i,j\}\,:\,S\cup\{i\}\in\mathcal{M},\,|S|=k\right\}=
  \left\{S\subseteq N\backslash\{i,j\}\,:\,S\cup\{j\}\in\mathcal{M},\,|S|=k\right\}.
$$   
With this we have $\sum_{h=1}^{l} \mathcal{M}_i(h)=\sum_{h=1}^{l} \mathcal{M}_j(h)$ for all $l\le k+1$. 
Now let $S\subseteq N\backslash\{i,j\}$ be a coalition of cardinality $k+1$. If $S\cup\{j\}\in \mathcal{M}$, then 
$S\cup\{i\}\in\mathcal{W}$ and $S'\cup\{j\}\notin\mathcal{M}$ for all $S'\subsetneq S$. Due to the definition of $k$ 
we have $S'\cup\{i\}\notin\mathcal{M}$ for all $S'\subsetneq S$ and conclude $S\cup\{i\}\in\mathcal{M}$. Using the 
definition of $k$ again, we conclude $\sum_{h=1}^{k+2} \mathcal{M}_i(h)=\sum_{h=1}^{k+2} \mathcal{M}_j(h)$ or $k=m-2$, 
i.e., we must have $i \succsim j$.

Since $\PGI_i=\sum_{l=1}^n |\mathcal{M}_i(l)|\ge \sum_{l=1}^n |\mathcal{M}_j(l)|=\PGI_j$ the statement is true for the
public Good index.

Since $ \DP_i = \dfrac 1l \sum_{l=1}^n |\mathcal M_i(l)|$ we have the decomposition
\begin{equation}
\label{E:difi-difj}
\begin{array}{rcl}
 \DP_i-\DP_j & = & \left( 1-\frac 12 \right) \left( \sum_{l=1}^1 |\mathcal M_i(l)| - \sum_{l=1}^1 |\mathcal M_j(l)| \right)  + \\
  &  & \left( \frac 12-\frac 13 \right) \left( \sum_{l=1}^2 |\mathcal M_i(l)| - \sum_{h=1}^2 |\mathcal M_j(l)| \right)  + \\
  &  & \dots  + \\
  &  & \left(\frac 1{n-1}-\frac 1n \right) \left( \sum_{l=1}^{n-1} |\mathcal M_i(l)| - \sum_{h=1}^{n-1} |\mathcal M_j(l)| \right)  + \\
  &  & \frac 1n  \left( \sum_{h=1}^{n} |\mathcal M_i(l)| - \sum_{h=1}^{n} |\mathcal M_j(l)| \right).
\end{array}
\end{equation}
Since $v$ is flat, each addend is non-negative, i.e., $\DP_i\ge \DP_j$. 

\end{Proof}

We remark that $i \sqsupset j$, $i \not\sqsupset j$ even implies $\DP_i > \DP_j$. Note further that uniform games are 
particular cases of flat games. 

\begin{example}
Let $v$ be the weighted game uniquely characterized by $N=\{1,2,3\}$ and $\mathcal M = \{ \{1\}, \{2,3\} \}$. This is a flat 
game with $1 \sqsupset 2$, $2 \not\sqsupset 1$, and $2 \square 3$. The (normalized) Public Good index is given by $\frac{1}{3}\cdot (1,1,1)$, 
i.e., player~$1$ and player~$2$ obtain the same value while not being equivalent. The (normalized) Deegan-Packel index 
is given by $\frac{1}{4}\cdot(2,1,1)$.  
\end{example}

\begin{corollary} 
  The Public Good and Deegan-Packel indices satisfy local monotonicity for weighted flat games.
\end{corollary}



\section{Convex combinations of power indices and the cost of local monotonicity}
\label{sec_convex_combinations}
\noindent
As mentioned in the introduction the aim of this paper is to study power indices arising as
a convex combination of a given (finite) collection of power indices. To this end let
$\mathcal{P}$ be a finite set of power indices, e.g.\ $\mathcal{P}=\left\{\BZ,
\PGI\right\}$ or $\mathcal{P}=\left\{\BZ,\PGI,\Shift\right\}$, which we will use later on. Given a set
$\mathcal{P}=\left\{P^1,\dots,P^r\right\}$ of power indices we consider the convex combinations
$$
  P^{\alpha,\mathcal{P}}=\sum_{i=1}^r \alpha_i\cdot P^i,
$$
where $\alpha=(\alpha_1,\dots,\alpha_r)\in [0,1]^r$ and $\sum_{i=1}^r \alpha_i=1$, that is $\alpha$ belongs to the $r$-dimensional simplex. For brevity we write $\alpha\in \mathbb{S}^r$. The power indices $P^i$ are defined
on possibly different classes $\mathfrak{G}_i$ and we set $\mathfrak{G}=\cap_{i=1}^r \mathfrak{G}_i$.
Obviously $P^{\alpha,\mathcal{P}}$ is a power index on $\mathfrak{G}$ too, i.e., it maps games in $\mathfrak{G}$
to a vector in $\mathbb{R}^n$. In the remaining part of the paper we will not explicitly mention the underlying
classes $\mathfrak{G}_i$ and $\mathfrak{G}$ of games.

Convex combinations of power indices have the nice feature that they
preserve the properties for power indices defined in Section~\ref{sec_notation}.

\begin{lemma}
  \label{lemma_properties}
  Let $\mathcal{P}=\left\{P^1,\dots,P^r\right\}$ be a collection of power indices such that $P^j$
  \begin{enumerate}
    \item[(1)] satisfies the null player property,
    \item[(2)] is symmetric,
    \item[(3)] is strictly positive,
    \item[(4)] is efficient,
    \item[(5)] has the dominance property,
    \item[(6)] is local monotonic, or
    \item[(7)] is invariant for nulls
  \end{enumerate}
  for all $1\le j\le r$, then $P^{\alpha,\mathcal{P}}$, where $\alpha\in\mathbb{S}^r$, has the same property.
\end{lemma}
\begin{Proof}
  For properties (1)-(3) the statement directly follows from the definition. For property~(4) we have
  $$
    \sum_{i=1}^n P_i^j(v)=1
  $$
  for all $1\le j\le r$. With this we conclude
  $$
    \sum_{i=1}^n P^{\alpha,\mathcal{P}}_i(v)=\sum_{i=1}^n\sum_{j=1}^r \alpha_j\cdot P_i^j(v)=
    \sum_{j=1}^r \alpha_j\cdot \sum_{i=1}^n P_i^j(v)=\sum_{j=1}^r \alpha_j=1,
  $$
  using the fact that the $\alpha_j$ sum up to one.

  For properties (5) and (6) we consider players $i$ and $h$ such that $P_i^j(v)\ge P_h^j(v)$ for all $1\le j\le r$.
  With this we have
  $$
    P^{\alpha,\mathcal{P}}_i(v)=\sum_{j=1}^r \alpha_j\cdot P_i^j(v)\ge
    \sum_{j=1}^r \alpha_j\cdot P_h^j(v)= P^{\alpha,\mathcal{P}}_h(v),
  $$
  since the $\alpha_j$ are non-negative.

  If the $P^j$ are invariant for nulls, then they have to satisfy the null player property. From (1) we deduce that
  $P^{\alpha,\mathcal{P}}$ also satisfies the null player property. Thus it suffices to prove
  $P^{\alpha,\mathcal{P}}_i(v)=P^{\alpha,\mathcal{P}}_i(v')$ for every player $i$ and every game $v'$ arising from $v$
  by deleting an arbitrary null player. We have
  $$
    P^{\alpha,\mathcal{P}}_i(v)=\sum_{j=1}^r \alpha_j\cdot P_i^j(v)=
    \sum_{j=1}^r \alpha_j\cdot P_i^j(v')= P^{\alpha,\mathcal{P}}_i(v'),
  $$
  so that the statement is also true for property~(7).
\end{Proof}

In this paper we are especially interested in the case where not all but at least one power index of a collection
$\left\{P^1,\dots,P^r\right\}$ satisfies local monotonicity. W.l.o.g.\ we assume $r \ge 2$ and that $P^1$ satisfies 
local monotonicity while the other indices might or might not satisfy LM. The convex combinations $P^{\alpha,\mathcal{P}}$ 
may or may not satisfy LM, depending on $\alpha$. At least for $\alpha=(1,0,\dots,0)$ LM is satisfied.

\begin{definition}
  \label{def_feasible_set}
  Let $\mathfrak{G}\subseteq\mathfrak{W}$ be a class of weighted games, $n\in\mathbb{N}_{>0}$ and $P=\left\{P^1,\dots,P^r\right\}$
  be a collection of power indices on $\mathfrak{G}$ such that $P^1$ satisfies local monotonicity. By
  $\mathbb{P}^{\mathcal{P}}_{\operatorname{LM}}(n,\mathfrak{G})$ we denote the set of
  $\alpha\in\mathbb{S}^r$ such that $P^{\alpha,\mathcal{P}}$ satisfies LM on the set of games of $\mathfrak{G}$
  consisting of $n$ players.
\end{definition}

\begin{lemma}
  \label{lemma_polyhedron}
  Given the requirements of Definition~\ref{def_feasible_set}, the set
  $\mathbb{P}^{\mathcal{P}}_{\operatorname{LM}}(n,\mathfrak{G})$ is a non-empty (bounded) polyhedron.
\end{lemma}
\begin{Proof}
  Obviously, we have $\mathbb{P}^{\mathcal{P}}_{\operatorname{LM}}(n,\mathfrak{G})\subseteq \mathbb{S}^r$. Given a game
  $v\in \mathfrak{G}$ and two players $1\le i,j\le n$ with $i\sqsupset j$, we have
  $P^{\alpha,\mathcal{P}}_i(v)\ge P^{\alpha,\mathcal{P}}_j(v)$ if and only if
  $P^{\alpha,\mathcal{P}}_i(v)-P^{\alpha,\mathcal{P}}_j(v)\ge 0$, which is equivalent to
  \begin{equation}
    \label{ie_feasible}
    \sum_{h=1}^r \alpha_h\cdot\underset{\in [0,1]}{\underbrace{\left(P^h_i(v)-P^h_j(v)\right)}} \ge 0.
  \end{equation}
  Thus $\mathbb{P}^{\mathcal{P}}_{\operatorname{LM}}(n,\mathfrak{G})$ is given as the intersection of $\mathbb{S}^r$ and
  the half-spaces (and possibly $\mathbb{R}^r$ for trivial inequalities $0\ge 0$) described by Inequality~(\ref{ie_feasible})
  for all $v\in \mathfrak{G}$ consisting of $n$ players and all
  $1\le i,j\le n$ with $i\sqsupset j$. We have $\mathfrak{G}\subseteq\mathfrak{W}\subseteq\mathfrak{S}$, so that
  the number of $n$-player games of $\mathfrak{G}$ is upper bounded by the number of simple games consisting of
  $n$ players. Since each simple game is uniquely characterized by its set of winning coalitions the number of
  simple games with $n$ players is at most $2^{2^n}$\!\!, i.e., finite. Thus we have a finite intersection of half-spaces
  and a polyhedron, which is a, possibly empty, polyhedron. It remains to remark that $(1,0,\dots,0)\in
  \mathbb{P}^{\mathcal{P}}_{\operatorname{LM}}(n,\mathfrak{G})$ to conclude the non-emptiness.
\end{Proof}

\begin{example}
\label{ex_1}
We consider $\mathcal{P}=\Big\{\overset{=:P^1}{\overbrace{\BZ}},\overset{=:P^2}{\overbrace{\PGI}},
\overset{=:P^3}{\overbrace{\Shift}}\Big\}$, $\mathfrak{G}=\mathfrak{W}$, and $n=7$, i.e., the class 
of weighted games with $7$ players. For $v=[2;2,1,1,1,1,1,1]$ we have $\BZ(v)=(7,5,\dots,5)$ and
$\PGI(v)=\Shift(v)=(1,5,\dots,5)$. The corresponding inequality~(\ref{ie_feasible}) for game $v$ and
players $1,2$ is given by $\alpha_1\cdot(7-5)+\alpha_2\cdot(1-5)+\alpha_3\cdot(1-5)\ge 0$. Inserting
$\alpha_2+\alpha_3=1-\alpha_1$ yields $2\alpha_1-4(1-\alpha_1)\ge 0$, which is equivalent to $\alpha_1\ge \frac{2}{3}$.
Thus for all $\alpha$ with $\alpha_1<\frac{2}{3}$ the convex combination $P^{\alpha,\mathcal{P}}$ does not satisfy
local monotonicity. This means that the weight of the Banzhaf index in a convex combination satisfying LM must be
at least $\frac{2}{3}$, i.e., closer to the Banzhaf index than to the two other indices, on the set of weighted games
with $7$ players.
\end{example}

\begin{example}
\label{ex_2}
Again we consider $\mathcal{P}=\Big\{\overset{=:P^1}{\overbrace{\BZ}},\overset{=:P^2}{\overbrace{\PGI}},
\overset{=:P^3}{\overbrace{\Shift}}\Big\}$, $\mathfrak{G}=\mathfrak{W}$, and $n=7$. For $v=[14;9,8,5,2,2,2,2]$
we have $\BZ(v)=(33,31,21,7,7,7,7)$, $\PGI(v)=(6,9,5,7,7,7,7)$, and $\Shift(v)=(1,8,5,4,4,4,4)$. The corresponding
inequality~(\ref{ie_feasible}) for the game $v$ and
players $1,2$ is given by $\alpha_1\cdot(33-31)+\alpha_2\cdot(6-9)+\alpha_3\cdot(1-8)\ge 0$. Inserting
$\alpha_1=1-\alpha_2+\alpha_3$ yields $\alpha_2\le \frac{2}{5}-\frac{9}{5}\cdot \alpha_3$ after a short calculation.
Thus for all $\alpha$ with $\alpha_2> \frac{2}{5}-\frac{9}{5}\cdot \alpha_3$ the convex combination $P^{\alpha,\mathcal{P}}$
does not satisfy local monotonicity.
\end{example}

\begin{figure}[htp]
\begin{center}
  \includegraphics{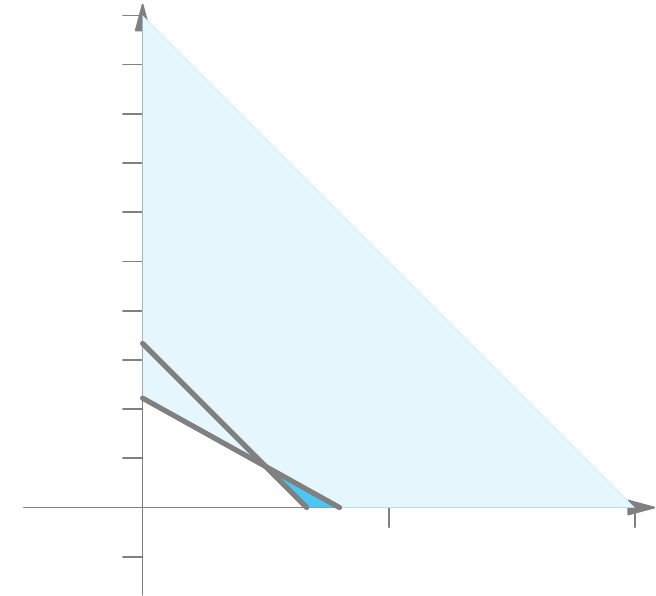}
\caption{Regions of $P^{\alpha,\left\{\BZ,\PGI,\Shift\right\}}$ which do not satisfy LM for weighted games with $n=7$ players.}
\label{fig_non_compatible_n_7}
\end{center}
\end{figure}

In Figure~\ref{fig_non_compatible_n_7} we have depicted the two regions, where $P^{\alpha,\mathcal{P}}$ does not satisfy
LM according to the weighted games of Example~\ref{ex_1} and Example~\ref{ex_2}. The weight region, including its border,
is a superset of $\mathbb{P}^{\mathcal{P}}_{\operatorname{LM}}(7,\mathfrak{W})$. Later on it will turn out it indeed coincides
with $\mathbb{P}^{\mathcal{P}}_{\operatorname{LM}}(7,\mathfrak{W})$.

Focusing on the necessary impact of $P^1$ over the rest of the $P^j$'s, we define the cost of local 
monotonicity $c_{\mathcal{P}}(n,\mathfrak{G})$ as the smallest value $\beta$ such that $P^{\alpha,\mathcal{P}}$ 
satisfies LM on $\mathfrak{G}$ for all $\alpha\in\mathbb{S}^r$ with $\alpha_1\ge \beta$.

\begin{definition}
  \label{def_cost_of_local_monotonicity}
  Given the requirements of Definition~\ref{def_feasible_set}, the cost of local monotonicity is defined as
  $$
    c_{\mathcal{P}}(n,\mathfrak{G})=\inf\left\{\beta\in [0,1] \mid P^{\alpha,\mathcal{P}}\text{satisfies LM on}\ (n,\mathfrak{G}),\,
    \ \ \forall\alpha\in\mathbb{S}^r\,:\,\alpha_1\ge \beta \right\}
  $$
  for $r>1$, where $(n,\mathfrak{G})$ stands for the class of games $\mathfrak{G}$ with at most $n$ voters, and 
  $c_{\mathcal{P}}(n,\mathfrak{G})=0$ for $r=1$.
\end{definition}

Some examples may be derived from the previous section. For $\mathfrak{G}=\mathfrak{U}$, or more generally  
$\mathfrak{G}=\mathfrak{F}$, and $\mathcal P = \{Jo,DP\}$ we have  $c_{\mathcal{P}}(n,\mathfrak{G})=0$ for all $n\in\mathbb{N}$, 
since the $DP$ index is local monotonic. Similarly, for $\mathfrak{G}=\mathfrak{U}$, or more generally  $\mathfrak{G}=\mathfrak{F}$, 
and $\mathcal P = \{Bz,PGI\}$ we have $c_{\mathcal{P}}(n,\mathfrak{G})=0$ for all $n\in\mathbb{N}$.

From Example~\ref{ex_1} we conclude
$c_{\left\{\BZ,\PGI,\Shift\right\}}(7,\mathfrak{W})\ge \frac{2}{3}$. The game from
Example~\ref{ex_2} gives the tighter inequality $c_{\left\{\BZ,\PGI,\Shift\right\}}(7,\mathfrak{W})\ge \frac{7}{9}$. Later
on it will turn out that we can replace the infimum in Definition~\ref{def_cost_of_local_monotonicity} by a minimum.
Next we remark that dropping some of the power indices that do no satisfy LM does not increase the cost of local monotonicity:

\begin{lemma}
  \label{lemma_cost_subset}
  Given the requirements of Definition~\ref{def_feasible_set}, let $P^1\in\mathcal{P}'\subseteq \mathcal{P}$
  be a subset, then we have $c_{\mathcal{P}'}(n,\mathfrak{G})\le c_{\mathcal{P}}(n,\mathfrak{G})$
  for all $n\in\mathbb{N}_{>0}$.
\end{lemma}
\begin{Proof}
  For $\left|\mathcal{P}'\right|=1$ the statement follows from $c_{\mathcal{P}}(n,\mathfrak{G})\ge 0$. For
  $\left|\mathcal{P}'\right|>1$ we can embed the elements of $\mathbb{S}^{\left|\mathcal{P}'\right|}$ in
  $\mathbb{S}^{\left|\mathcal{P}\right|}$ by choosing zero for the missing indices.
\end{Proof}

\begin{observation}
  Given the requirements of Definition~\ref{def_feasible_set}, we have
  $c_{\mathcal{P}}(n,\mathfrak{G}')\le c_{\mathcal{P}}(n,\mathfrak{G})$ for all $\mathfrak{G}'\subseteq \mathfrak{G}$.
\end{observation}

Under slight technical assumptions on the set of power indices and on $\mathfrak{G}$ we have monotonicity in the number of players:

\begin{lemma}
  Given the requirements of Definition~\ref{def_feasible_set}, assume that all $P^j$'s are invariant for nulls and
  that $\mathfrak{G}$ is closed under the addition of null players. Then we have
  $c_{\mathcal{P}}(n,\mathfrak{G})\le c_{\mathcal{P}}(n+1,\mathfrak{G})$ for all $n\in\mathbb{N}_{>0}$.
\end{lemma}
\begin{Proof}
  Given an $n$-player game $v\in \mathfrak{G}$ with $P^{\alpha,\mathcal{P}}(v)=(p_1,\dots,p_n)$, we can construct
  a game $v'\in \mathfrak{G}$ by adding a null player such that
  $P^{\alpha,\mathcal{P}}(v')=(p_1,\dots,p_n,0)$. Since $P^{\alpha,\mathcal{P}}$ satisfies LM for $v$
  if and only if $P^{\alpha,\mathcal{P}}$ satisfies LM for $v'$, the statement follows.
\end{Proof}

For collections of $r=2$ power indices not only the set $\mathbb{P}^{\mathcal{P}}_{\operatorname{LM}}(n,\mathfrak{G})$
is a polyhedron but also its complement $\mathbb{S}^r\backslash \mathbb{P}^{\mathcal{P}}_{\operatorname{LM}}(n,\mathfrak{G})$
since both sets are intervals. Thus we can replace the infimum in Definition~\ref{def_cost_of_local_monotonicity} by a minimum
for all cases where $r=2$. Next we will show that the lower bounds from Lemma~\ref{lemma_cost_subset} for all subsets
of cardinality two are sufficient to determine the cost of local monotonicity in general:

\begin{lemma}
  \label{lemma_cost_only_2}
  Given the requirements of Definition~\ref{def_feasible_set}, we have
  $$
    c_{\mathcal{P}}(n,\mathfrak{G})=\max \left\{c_{\left\{P^1,P^j\right\}}(n,\mathfrak{G})\mid 2\le j\le r \right\}.
  $$
\end{lemma}
\begin{Proof}
  From Lemma~\ref{lemma_cost_subset} we conclude $c_{\mathcal{P}}(n,\mathfrak{G})\ge\max \left\{c_{\left\{P^1,P^j\right\}}(n,\mathfrak{G})\mid 2\le j\le r \right\}$.
  W.l.o.g.\ we assume $c_{\left\{P^1,P^j\right\}}(n,\mathfrak{G})<1$ for all $2\le j\le r$.
  Let $e_i$ denote the $i$th unit vector. With this define $v_j=e_1\cdot c_{\left\{P^1,P^j\right\}}(n,\mathfrak{G})+e_j\cdot
  \left(1-c_{\left\{P^1,P^j\right\}}(n,\mathfrak{G})\right)$ for all $2\le j\le r$. We have already observed that for $r=2$
  the infimum in the definition of the cost of local monotonicity is indeed attained. Thus $P^{v_j,\mathcal{P}}$ satisfies
  LM. Since $\mathbb{P}^{\mathcal{P}}_{\operatorname{LM}}(n,\mathfrak{G})\subseteq \mathbb{S}^r$ is convex,
  the $n-1$-dimensional simplex $\operatorname{conv}(e_1,v_2,\dots,v_r)=:\mathbb{F}$ is contained in
  $\mathbb{P}^{\mathcal{P}}_{\operatorname{LM}}(n,\mathfrak{G})$. Now let $\mathbb{A}$ be the closure of
  $\mathbb{S}^r\backslash\mathbb{F}$, which is a polyhedron too. Indeed, the vertices of $\mathbb{A}$ are given by
  $v_2,\dots, v_r$ and $e_2,\dots,e_r$.

  With this we have $c_{\mathcal{P}}(n,\mathfrak{G})\le \max\{a_1\mid (a_1,\dots,a_r)\in\mathbb{A}\}$, where the maximum
  is attained at one of the $2r-2$ vertices of $\mathbb{A}$. For the vertices $e_2,\dots,e_r$ the respective $a_1$-value
  is zero. By definition the $a_1$-value of $v_j$ is given by $c_{\left\{P^1,P^j\right\}}$ for all $2\le j\le r$.
  Thus, $c_{\mathcal{P}}(n,\mathfrak{G})\le\max \left\{c_{\left\{P^1,P^j\right\}}(n,\mathfrak{G})\mid 2\le j\le r \right\}$.
\end{Proof}

\begin{corollary}
  $$
    c_{\mathcal{P}}(n,\mathfrak{G})=\min\left\{\beta\in\mathbb{R}_{\ge 0}\mid P^{\alpha,\mathcal{P}}\text{ satisfies LM on }\mathfrak{G}\,
    \forall\alpha\in\mathbb{S}^r\,:\,\alpha_1\ge \beta\right\}
  $$
\end{corollary}

So, in order to determine $c_{\mathcal{P}}(n,\mathfrak{G})$ it suffices to determine $c_{\left\{P^1,P^h\right\}}(n,\mathfrak{G})$
for all $2\le h\le r$. Given a game $v\in\mathfrak{G}$ and two players $i,j\in N$ with $i\sqsupset j$, we can solve
Inequality~(\ref{ie_feasible}) for $\alpha_1$ using $\alpha_1+\alpha_h=1$ If $P^h$ violates LM for $v$ and
players $i,j$, we obtain an inequality of the form $\alpha_1\ge u$ and can conclude $c_{\left\{P^1,P^h\right\}}(n,\mathfrak{G})\ge u$.
>From Example~\ref{ex_1} we conclude $c_{\left(\BZ,\PGI\right)}(7,\mathfrak{W})\ge\frac{2}{3}$ and from Example~\ref{ex_2}
we conclude $c_{\left(\BZ,\PGI\right)}(7,\mathfrak{W})\ge\frac{7}{9}$.

In order to obtain tight bounds for the cost of local monotonicity, we may simply loop over all possible choices
of $v$, $i$, and $j$. At the very least, we can partially restrict the number of choices as follows: Assuming
$1\sqsupset \dots \sqsupset n$ the requirement $i\sqsupset j$ is equivalent to $i\le j$. We remark that if $P^h$ violates
LM for $v$ and players $i,j$ and there exists another player $i<i'<j$, then LM is violated for $v$ and at least one of the pairs $i,i'$
or $i',j$ of players. Thus, we can restrict our considerations on pairs of players of the form $i,i+1$, where $1\le i<n$.

\begin{lemma}
  \label{lemma_enumeration}
  Given the requirements of Definition~\ref{def_feasible_set} with $r=2$, we have
  $$
    c_{\left(P^1,P^2\right)}(n,\mathfrak{G})=\max\left\{l_i(v)\mid v\in\mathfrak{G},\,1\le i\le n-1\right\},
  $$
  where
  $$l_i(v):=\left(P_{i}^2(v)-P_{i+1}^2(v)\right)/\left(P_{i+1}^1(v)-P_{i}^1(v)+P_{i}^2(v)-P_{i+1}^2(v)\right)$$ if
  $P_{i+1}^1(v)-P_{i}^1(v)+P_{i}^2(v)-P_{i+1}^2(v)>0$ and $l_i(v):=0$ otherwise.
\end{lemma}

The big drawback of this exact approach is the usually large size of the set of $n$-player games of $\mathfrak{G}$.
Both sets of $n$-player complete simple or weighted games grow faster than exponential. The exact numbers have been determined 
up to $n=9$ only, see e.g.\ \cite{krohn1995} for the numbers of complete and weighted games up to $n=8$, \cite{freixas2010without} 
for the number of complete games for $n=9$, and  \cite{kurz2012minimum,kurz2013dedekind} for the number of weighted games for $n=9$. 
For $n=9$ there are
$284\,432\,730\,174$ complete simple and $993\,061\,482$ weighted games.\footnote{which had to be
slightly corrected recently \cite{cutting}.} Thus, using Lemma~\ref{lemma_enumeration}
becomes computationally infeasible for $n>9$. So we propose an integer linear programming formulation in the next section.

\section{An integer linear programming formulation}
\label{sec_ILP}

\noindent
Whenever one is interested in complete simple games or weighted games, which are extremal with respect to a certain
criterion, exhaustive enumeration is not a feasible option for $n>9$ players, see the enumeration results stated at the
end of the previous section. An alternative is to specify the set games indirectly by binary variables and linear
inequalities. If the extremality criterion can be also formulated using integer variables and linear constraints, then
integer linear programming techniques can be applied. In the context of cooperative games this approach was introduced
in \cite{kurz2012inverse} and also applied in this context in e.g.\ \cite{freixas2011alpha,inverse,kurz2012heuristic}.

For completeness, we briefly repeat the ILP formulation of a game $v$. Since $v$ is uniquely characterized by its values
$v(S)$ for all coalitions $S\in 2^N$, we introduce binary variables $x_S\in\{0,1\}$ for all $S\in 2^N$. The conditions
for a simple game can be stated as $x_{\emptyset}=0$, $x_N=1$, and $x_S\le x_T$ for all $S\subseteq T\subseteq N$.
We remark that for the later set of inequalities it suffices to consider the pairs of coalitions where $|T|=|S|+1$. Complete
simple games can be modeled by additionally requiring $x_S\le x_T$ for all pairs of coalitions with
$t_i\sqsupset s_i$ for $1\le i\le m$, where $S=\{s_1,\dots,s_m\}$ and $T=\{t_1,\dots,t_m\}$.

In order to restrict $v$ to weighted games we additionally have to introduce weights $w_i\ge 0$ and a quota $q>0$, where
we assume that the weight of each winning coalition is larger than the weight of each losing coalition by at
least one. (We may simply use integer weights, which could
result in harder problems for the ILP solver.) To interlink the $x_S$ with the $w_i$ and $q$ we use
\begin{eqnarray*}
  q-(1-x_S)\cdot M-\sum\limits_{i\in S} w_i&\le& 0\quad\quad\forall S\in 2^N\text{ and}\label{iq_big_M_1}\\
  -x_S\cdot M+\sum\limits_{i\in S} w_i&\le& q-1\quad\quad\forall S\in 2^N\label{iq_big_M_2},
\end{eqnarray*}
where $M$ is a suitably large constant fulfilling $M-1\ge\sum\limits_{i=1}^n w_i$. (We may choose
$M=4n\left(\frac{n+1}{4}\right)^{(n+1)/2}$, see \cite[Theorem 9.3.2.1]{0243.94014}.)

The restrictions to proper games can be formulated via $x_S+x_{N\backslash S}\le 1$ for all $S\subseteq N$ with
$|S|\le\frac{n}{2}$. Similarly we can restrict to strong games by requesting $x_S+x_{N\backslash S}\ge 1$.  For
constant sum games we need $x_S+x_{N\backslash S}= 1$.

In order to compute the power distribution of $\BZ$, $\PGI$, and $\Shift$ from the $x_S$, we introduce further binary variables,
cf.~\cite{inverse}. For $i\in N$ and $S\in 2^N$ we set $y_{i,S}=1$ if and only if coalition $S$ is a swing for player~$i$ and $y_{i,S}=0$
otherwise. This can be ensured by requesting $y_{i,S}=0$ for $i\notin S$ and $y_{i,S}=x_S-x_{S\backslash\{i\}}$ otherwise. Similarly
we introduce $z_S\in\{0,1\}$, where $z_S=1$ if and only if $S$ is a minimal winning coalition. This condition can be linearly reformulated
as
\begin{eqnarray*}
  z_S-x_S &\le& 0 \quad\quad\forall S\in 2^N,\\
  z_S+x_{S\backslash\{i\}} &\le& 1 \quad\quad\forall S\in 2^N,\, i\in S,\text{ and}\\
    z_S-x_S+\sum_{i\in S} x_{S\backslash\{i\}} &\ge& 0 \quad\quad\forall S\in 2^N.
\end{eqnarray*}
In order to identify shift-minimal winning coalitions, we introduce binary variables $u_S\in\{0,1\}$ for all $S\in 2^N$. In order
to state characterizing linear constraints we additionally need binary variables $t_i\in\{0,1\}$ for all $1\le i\le n-1$,
which are equal to $0$ if and only if players $i$ and $i+1$ are of the same type, i.e., $i\,\square\, i+1$. This equivalence can be
ensured by requesting
\begin{eqnarray*}
  x_{S\cup\{i\}}-x_{S\cup\{i+1\}}-t_i &\le& 0\quad\forall 1\le i\le n-1,\, S\subseteq N\backslash\{i,i+1\}\text{ and}\\
  -t_i+\sum_{S\subseteq N\backslash\{i,i+1\}} x_{S\cup\{i\}}-x_{S\cup\{i+1\}} &\ge& 0\quad\forall 1\le i\le n-1.
\end{eqnarray*}
If $t_i=0$, i.e., $i\,\square\,i+1$, then $S\cup\{i\}$ is a shift-minimal winning coalition if and only if
$S\cup\{i+1\}$ is a shift-minimal winning coalition, where $S\subseteq N\backslash\{i,i+1\}$ and $1\le i\le n-1$. This
conditioned equivalence can be expressed as
\begin{eqnarray*}
  u_{S\cup\{i\}}\ge  u_{S\cup\{i+1\}}-t_i\quad\text{and}\quad u_{S\cup\{i+1\}}\ge  u_{S\cup\{i\}}-t_i.
\end{eqnarray*}
So in the following we can restrict our considerations on coalitions $S$ such that for each $i\in S$ we either
have $i=n$, $i+1\in S$ or $i \sqsupset i+1$.

Since each shift-minimal winning coalition has to be a minimal winning coalition, we require $u_S\le z_S$ for all
$S\in 2^N$. The other possibility disqualifying a coalition $S$ from being a shift-minimal winning coalition
is the existence of player $i\in S\backslash\{n\}$ with $i+1\notin S$ and $i\sqsupset i+1$ such that $S\cup\{i+1\}\backslash\{i\}$
is winning. So we require
$$
  u_S\le 1+x_S-x_{S\cup\{i+1\}\backslash\{i\}}-t_i
$$
for all $S\subseteq N$, $n\neq i\in S$ and $i+1\notin S$. Since $x_S\ge x_{S\cup\{i+1\}\backslash\{i\}}$ the right
hand side is at least zero. So let us assume $t_i=1$. Since $u_S\le z_S\le x_S$ it suffices to consider the cases where $x_S=1$.
If $x_{S\cup\{i+1\}\backslash\{i\}}=1$ then the stated inequality is trivially true. Just in the single case, where
$x_S=1$, $x_{S\cup\{i+1\}\backslash\{i\}}=0$, and $t_i=1$, it implies $u_S=0$.

By now we can guarantee that $u_S=0$ if $S$ is not a shift-minimal
winning coalition. However $u_S=0$ is still feasible for shift-minimal winning coalitions.
So, we additionally require
$$
  u_S-x_S+\sum_{i\in N\backslash\{n\}:i\in S,i+1\notin S}x_{S\cup\{i+1\}\backslash\{i\}}\ge 0
$$
for all $S\subseteq N\backslash{n}$ and
$$
  u_S-x_S+x_{S\backslash\{n\}}+\sum_{i\in N\backslash\{n\}:i\in S,i+1\notin S}x_{S\cup\{i+1\}\backslash\{i\}}\ge 0
$$
for all $\{n\}\subseteq S \subseteq N$.
If $x_S=0$ or one of the $x_{S\cup\{i+1\}\backslash\{i\}}=1$, then the proposed inequality is trivially satisfied.
So we assume otherwise. In the cases where $n\in S$ we can similarly assume $x_{S\backslash\{n\}}=0$. In this remaining
case we have the implication $u_S=1$, which is correct since no certificate for not being a shift-minimal winning coalition
exists, so that $S$ is a shift-minimal winning coalition.

Finally consider the case where $S$ indeed is a shift-minimal winning coalition. Thus $x_S=1$. Remember that we need the implication
$u_S\ge 1$ just for the coalitions $S$, where for each player $i\in S$ we either have $i=n$, $i+1\in S$, or $i \sqsupset i+1$.
So we can assume $i \sqsupset i+1$ for all indices $i$ in the summation. Thus $x_{S\cup\{i+1\}\backslash\{i\}}=0$. If $n\in S$,
then also $x_{S\backslash\{n\}}=0$ since $S$ is a minimal winning coalition.

\medskip

Having these variables at hand, we can easily compute the following power indices:
\begin{eqnarray*}
  \BZ_i(v) &=& \sum_{S\subseteq N} y_{i,S},\\
  \PGI_i(v) &=& \sum_{\{i\}\subseteq S\subseteq N} z_{S},\\
  \Shift_i(v) &=& \sum_{\{i\}\subseteq S\subseteq N} u_{S},\\
  \DP_i(v) &=& \sum_{\{i\}\subseteq S\subseteq N} \frac{1}{|S|}\cdot z_{S}, \text{ and }\\
  \SDP_i(v) &=& \sum_{\{i\}\subseteq S\subseteq N} \frac{1}{|S|}\cdot u_{S}.\\
\end{eqnarray*}

For the Johnston index we have to take care that only the swing players obtain an equal share for each coalition.
To this end we introduce the continuous variables $b_{i,S}\in\mathbb{R}_{\ge 0}$ for all $i\in N$ and and all $S\in 2^N$,
cf.~\cite{inverse}:

\begin{eqnarray*}
  b_{i,S}&\le& y_{i,S}\quad\forall S\in 2^N,\, i\in N\\
  b_{i,S}-b_{j_S} &\ge& y_{i,S}+y_{j,S}-2\quad \forall S\in 2^N,\, i,j\in N\\
  \sum_{i=1}^n b_{i,S} &\le& 1\quad \forall S\in 2^N\\
  \sum_{j=1}^n b_{i,S} &\ge& y_{i,S}\quad    \forall S\in 2^N,\, i\in N.
\end{eqnarray*}

Given an arbitrary coalition $S\subseteq N$, we can easily check that $\sum_{i=1}^n b_{i,S}=0$ if
$S$ is not a swing coalition for any player $i\in N$ and $\sum_{i=1}^n b_{i,S}=1$ otherwise. In the
later case we have $b_{i,S}=0$ whenever player $i$ is not a swing for coalition $S$. The second set
of inequalities guarantees $b_{i,S}=b_{j,S}$ whenever both $i$ and $j$ are swings for coalition $S$. So we
can state
\begin{eqnarray*}
  \J_i(v)&=& \sum_{S\subseteq N} b_{i,S}.
\end{eqnarray*}

As a target we maximize
$$
  P^{\alpha,\mathcal{P}}_{i+1}(v)-P^{\alpha,\mathcal{P}}_{i}(v)=\sum_{h=1}^{r} \alpha_h P^h_{i+1}(v)-\alpha_hP^h_{i}(v),
$$
where $1\le i\le n-1$ has to be specified as a parameter. By looping over all possible values of $i$ we
can decide whether $P^{\alpha,\mathcal{P}}$ satisfies LM for a given $\alpha\in\mathbb{S}^r$.

\begin{lemma}
  \label{lemma_feasibility}
  For $\mathcal{P}\subseteq\{\BZ,\PGI,\Shift,\J,\DP,\SDP\}$, $r:=|\mathcal{P}|$, and $\alpha\in \mathbb{S}^r$, one
  can decide $\alpha\in \mathbb{P}^{\mathcal{P}}_{\text{LM}}(n,\mathfrak{G})$ if incidence vectors of the $n$-player games
  in $\mathfrak{G}$ form a polyhedron.
\end{lemma}

To be more precise, we have explicitly stated ILP formulations for the classes of games
$\mathfrak{G}\in\left\{\mathfrak{S},\mathfrak{C},\mathfrak{W},\mathfrak{S}^p,\mathfrak{C}^p,\mathfrak{W}^p,\mathfrak{S}^s,
\mathfrak{C}^s,\mathfrak{W}^s,\mathfrak{S}^c,\mathfrak{C}^c,\mathfrak{W}^c\right\}$. Having the binary variables $t_i$ at
hand restrictions on the number of equivalence classes of players or even the precise partition can be formulated easily.
ILP formulations for further power indices can be found in \cite{inverse}.

We assume that the algorithm behind Lemma~\ref{lemma_feasibility} gives either the answer yes, if all corresponding $n-1$
ILPs have an optimal target value of zero, or gives the answer no together with a game $v\in\mathfrak{G}$ and an index
$1\le i\le n-1$ such that $P^{\alpha,\mathcal{P}}$ violates LM for the game $v$ and players $i$, $i+1$. The following algorithm
computes the cost of local monotonicity for convex combinations of two power indices, i.e., $\mathcal{P}=\{P^1,P^h\}$:

\bigskip

\noindent
$\alpha_1=0$\\
$\alpha_h=1$\\
$go\_on=true$\\
\textbf{while} $go\_on=true$ \textbf{do}\\
\hspace*{5mm} $go\_on=false$\\
\hspace*{5mm} \textbf{for} $i$ \textbf{from} $1$ \textbf{to} $n-1$ \textbf{do}\\
\hspace*{10mm} \textbf{if} $\max \alpha_1P_{i+1}^1(v)-\alpha_1P_{i}^1(v)+\alpha_hP_{i+1}^h(v)-\alpha_hP_{i}^h(v)>0$ \textbf{then}\\
\hspace*{15mm} $go\_on=true$\\
\hspace*{15mm} choose $v^\star\in\arg \max \alpha_1P_{i+1}^1(v)-\alpha_1P_{i}^1(v)+\alpha_hP_{i+1}^h(v)-\alpha_hP_{i}^h(v)>0$\\
\hspace*{15mm} determine $\beta$ with $\beta\left(P_{i+1}^1(v^\star)-P_{i}^1(v^\star)\right)+(1-\beta)\left(P_{i+1}^h(v^\star)-P_{i}^h(v^\star)\right)=0$\\
\hspace*{15mm} $\alpha_1=\beta$\\
\hspace*{15mm} $\alpha_h=1-\beta$\\
\hspace*{10mm} \textbf{end if}\\
\hspace*{5mm} \textbf{end for}\\
\textbf{end while}

\section{Exact values and lower bounds for the cost of local monotonicity}
\label{sec_exact_values_and_bounds}
By considering parametric examples we can obtain general lower bounds for the cost of local monotonicity.

\begin{lemma}
  \label{lemma_lower_bound_PGI_weighted}
  For $\mathcal{P}=\left(\BZ,\PGI\right)$ and $n\ge 2$ we have
  $c_{\mathcal{P}}(n,\mathfrak{W})\ge \max\!\left(0,\frac{n-3}{n-1}\right)$.
\end{lemma}
\begin{Proof}
  Since $c_{\mathcal{P}}(n,\mathfrak{W})\ge 0$ by definition, it suffices to consider weighted games with $n\ge 4$ {\voter}s.
  For the weighted game $v=[2;2,1,\dots,1]$, with $n-1$ players of weight $1$ and one player of weight $2$, the minimal
  winning coalitions are given by $\{1\}$ and $\{i,j\}$, where $2\le i<j\le n$. Thus, we have $\PGI_1(v)=1$ and
  $\PGI_2(v)=n-2$. For player~$1$ the swing coalitions are given by $\{1\}$ and $\{1,i\}$ for
  all $2\le i\le n$. Given a player~$j\ge 2$ the swing coalitions for player~$j$ are given by $\{i,j\}$ for
  all $2\le i\le n$, with $i\neq j$. Thus, we have $\BZ_1(v)=n$ and $\BZ_2(v)=n-2$.

  For players~$1$, $2$ and game $v$ Inequality~(\ref{ie_feasible}) reads
  $$
    \alpha_1\cdot \BZ_1(v)+\alpha_2\cdot\PGI_1(v) \ge
    \alpha_1\cdot \BZ_2(v)+\alpha_2\cdot\PGI_2(v),
  $$
  which is equivalent to
  $$
    \alpha_1\cdot n+\alpha_2\cdot 1 \ge\alpha_1\cdot (n-2)+\alpha_2\cdot(n-2)
    \quad\Longleftrightarrow\quad
    \alpha_1\ge \frac{n-3}{n-1},
  $$
  since $\alpha_1+\alpha_2=1$.
\end{Proof}

\begin{corollary}
  \label{cor_lower_bound_Shift_weighted}
  For $\mathcal{P}=\left(\BZ,\Shift\right)$ and $n\ge 2$ we have
  $c_{\mathcal{P}}(n,\mathfrak{W})\ge \max\!\left(0,\frac{n-3}{n-1}\right)$.
\end{corollary}
\begin{Proof}
  Since all minimal winning coalitions in the example of the proof of Lemma~\ref{lemma_lower_bound_PGI_weighted}
  are also shift-minimal winning, we can apply the same proof for the Shift index.
\end{Proof}

So, from Lemma~\ref{lemma_lower_bound_PGI_weighted}, Corollary~\ref{cor_lower_bound_Shift_weighted}, and
Lemma~\ref{lemma_cost_only_2} we can conclude that the cost of local monotonicity
is at least $\frac{n-3}{n-1}$ for $\mathcal{P}=\left(\BZ,\PGI,\Shift\right)$ and all $n\ge 2$.

\begin{corollary}
  $$
    \lim_{n\to\infty} c_{\left(\BZ,\PGI\right)}(n,\mathfrak{W})=
    \lim_{n\to\infty} c_{\left(\BZ,\Shift\right)}(n,\mathfrak{W})=
    \lim_{n\to\infty} c_{\left(\BZ,\PGI,\Shift\right)}(n,\mathfrak{W})=1
  $$
\end{corollary}
In other words, the only convex combination of $\BZ$, $\PGI$, and $\Shift$ that is locally monotonic for all weighted
games is the raw Banzhaf index itself. For a finite number of players it may still be possible that is cost of
local monotonicity is strictly less than $1$.

Having the ILP approach from the previous section at hand we can also determine the exact value of the cost of
local monotonicity for a small number of {\voter}s. It turns out that the lower bound from Lemma~\ref{lemma_lower_bound_PGI_weighted}
is tight for all $n\le 10$. So, especially for $n\le 3$ we have a cost of local monotonicity of zero, which goes in line
with the fact that all weighted games with at most $3$ {\voter}s are locally monotonic. Although the ILP approach
can move the computational limit of exhaustive enumeration a bit, it is so far limited to $n\le 10$. Since we were not
able to find worser examples, we state:

\begin{conjecture}
  \label{conj_BZ_PGI_weighted}
  For $\mathcal{P}=\left(\BZ,\PGI\right)$ and $n\ge 2$ we have $c_{\mathcal{P}}(n,\mathfrak{W})=\max\!\left(0,\frac{n-3}{n-1}\right)$.
\end{conjecture}

We remark that the extremal examples are not unique, e.g., we have the weighted games
$[4;4,3,2,2,1,1]$, $[10;10,8,5,4,4,3,3,2]$ and $[7;7,6,6,6,5,3,3,3,2,1]$ also meeting the bound from
Lemma~\ref{lemma_lower_bound_PGI_weighted} for $n=6$, $n=8$, and $n=10$, respectively. The respective
Banzhaf scores are given by $(11,9,5,5,3,3)$, $(28,26,16,12,12,10,10,6)$, and $(24,22,22,22,20,12,12,12,8,6)$.
The respective PGI scores are given by $(1,4,3,3,3,3)$, $(1,6,11,9,9,8,8,6)$, and $(1,8,8,8,7,9,9,9,7,6)$.

For combinations of the Banzhaf score and the Shift score the lower bound of Corollary~\ref{cor_lower_bound_Shift_weighted}
is tight for $n\le 6$. For $7\le n\le 11$ we were able to computationally find worser examples.

\begin{lemma}
  \label{lemma_lower_bound_Shift_weighted_special}
  Let $\mathcal{P}=\left(\BZ,\Shift\right)$.
  \begin{enumerate}
    \item[(1)] For $n=7$ we have $c_{\mathcal{P}}(n,\mathfrak{W})= \frac{7}{9}\approx 0.77777$.
    \item[(2)] For $n=8$ we have $c_{\mathcal{P}}(n,\mathfrak{W})= \frac{7}{8}=0.875$.
    \item[(3)] For $n=9$ we have $c_{\mathcal{P}}(n,\mathfrak{W})= \frac{25}{27}\approx 0.92593$.
    \item[(4)] For $n=10$ we have $c_{\mathcal{P}}(n,\mathfrak{W})= \frac{51}{53}\approx 0.96226$.
    \item[(5)] For $n=11$ we have $c_{\mathcal{P}}(n,\mathfrak{W})= \frac{97}{99}\approx 0.97980$.
    \item[(6)] For $n=12$ we have $c_{\mathcal{P}}(n,\mathfrak{W})\ge \frac{175}{177}\approx 0.98870$.
  \end{enumerate}
\end{lemma}
\begin{Proof}
  For the lower bounds we state an explicit weighted game and the Banzhaf and the Shift score for the first
  two {\voter}s:
  \begin{enumerate}
    \item [(1)]$[14;9,8,5,2,2,2,2]$, $\BZ=(33,31,\dots)$, $\Shift=(1,8,\dots)$;
    \item [(2)]$[16;11,10,5,2,2,2,2,2]$, $\BZ=(65,63,\dots)$, $\Shift=(1,15,\dots)$;
    \item [(3)]$[30;16,15,7,7,3,3,3,3,3]$, $\BZ=(129,127,\dots)$, $\Shift=(2,27,\dots)$,\\
               $[18;13,12,5,2,2,2,2,2,2]$, $\BZ=(129,127,\dots)$, $\Shift=(1,26,\dots)$;
    \item [(4)]$[33;19,18,7,7,3,3,3,3,3,3]$, $\BZ=(257,255,\dots)$, $\Shift=(2,53,\dots)$;
    \item [(5)]$[36;22,21,7,7,3,3,3,3,3,3,3]$, $\BZ=(513,511,\dots)$, $\Shift=(2,99,\dots)$.
    \item [(6)]$[56;29,28,9,9,9,4,4,4,4,4,4,4]$, $\BZ=(1025,1023,\dots)$, $\Shift=(2,177,\dots)$.
  \end{enumerate}
  For the upper bounds we have applied the ILP approach.
\end{Proof}

In some cases we have already stated different tight examples. We would highlight just another tight example, where
the bound is met between the second and the third {\voter}, for $n=9$: $[8;5,3,2,2,2,2,2,2,2]$ with Banzhaf score
$(85,43,41,41,41,41,41,41,41)$ and Shift score $(22,1,26,26,26,26,26,26,26)$.

We provide a general construction meeting the best known examples for all $n\ge 6$:

\begin{lemma}
  \label{lemma_general_construction}
  Let $k\ge 1$ be an integer, $m\in\{0,1,2\}$, $n=3k+3+m$, and
  $$
    v=\big[2t+m(k+1);t+1+m(k+1),t+m(k+1),\overset{k\text{ times}}{\overbrace{2k+3,\dots,2k+3}},
    \overset{2k+1+m\text{ times}}{\overbrace{k+1,\dots,k+1}}\big],
  $$
  where $t=2k^2+3k+1$. With this, $v$ is a weighted game consisting of $n$ players, $\BZ_1(v)=2^{n-2}+1$,
  $\BZ_2(v)=2^{n-2}-1$, $\Shift_1(v)=1$ for $k=1$, $\Shift_1(v)=2$ for $k>1$, and
  $$
    \Shift_2(v)=-1+\Shift_1(v)+\sum_{a=0}^{k}\sum_{b=\left\lceil\frac{t-a(2k+3)}{k+1}\right\rceil}
    {k \choose a}\cdot{{2k+1+m}\choose b}.
  $$
\end{lemma}
\begin{Proof}
  We can easily check that $v$ consists of $n=3k+3+m$ {\voter}s having $4$ different non-negative weights.
  For the ease of notation we denote coalition types as a $4$-tuple $(c_1,c_2,c_3,c_4)$ meaning
  a coalition having $0\le c_1\le 1$ {\voter}s of weight $t+1+m(k+1)$, $0\le c_2\le 1$ {\voter}s of weight
  $t+m(k+1)$, $0\le c_3\le k$ {\voter}s of weight $2k+3$, and $0\le c_4\le 2k+1+m$ {\voter}s of weight
  $k+1$. Due to symmetry it suffices to know the counts $c_1,\dots,c_4$ instead of the precise coalitions.

  Note that the sum of weights of the {\voter}s of weight $2k+3$ is given by $t-1$. The sum of weights of
  {\voter}s of weight $k+1$ is given by $t+m(k+1)$. Next we show that the {\voter}s of the $4$ weight types
  are non-equivalent. Since $(1,1,0,0)$ is winning and $(1,0,1,0)$ is losing, the {\voter} of weight $t+m(k+1)$
  is non-equivalent to {\voter}s of weight $2k+3$. Since $(1,0,k,0)$ and winning and $(0,1,k,0)$ is losing
  the first two {\voter}s are non-equivalent. Since $(1,0,k,0)$ and winning and $(1,0,0,k)$ is losing, also
  the {\voter}s of weight $2k+3$ are non-equivalent to {\voter}s of weight $k+1$. Due to the decreasing sequence
  of weights, we have four equivalence classes of {\voter}s coinciding with the sets of {\voter}s that have
  an equal weight.

  A coalition type $(1,1,a,b)$ corresponds to swing coalitions for {\voter}~$1$ if and only if $a(2k+3)+b(k+1)\le t-1$.
  Similarly, a coalition type $(1,0,a,b)$ corresponds to swing coalitions for {\voter}~$1$ if and only if $a(2k+3)+b(k+1)\ge t-1$.
  So we are interest in the number of cases where $a(2k+3)+b(k+1)= t-1$. Considering this equation modulo $k+1$
  yields $a\equiv -1\pmod k+1$, so that $a=k$, $b=0$ is the unique solution. Thus we have $\BZ_1(v)=2^{n-2}+1$.

  A coalition type $(1,1,a,b)$ corresponds to swing coalitions for {\voter}~$2$ if and only if $a(2k+3)+b(k+1)\le t-2$ and
  a coalition type $(0,1,a,b)$ corresponds to swing coalitions for {\voter}~$2$ if and only if $a(2k+3)+b(k+1)\ge t$.
  Thus we have $\BZ_1(v)=2^{n-2}-1$.

  The coalition $\{1,2\}$ is a minimal winning coalition in any case. We can easily check that it is
  shift-minimal winning if and only if $k>1$. Since the weight of $(1,0,k,0)$ exactly equals the quota, the corresponding
  unique coalition is shift-minimal in any case. Now assume that a coalition $S$ of type $(1,0,a,b)$ is shift-minimal
  winning. Since $S$ has to be winning, we have $a(2k+3)+b(k+1)\ge t-1$. Since $(0,1,a,b)$ has to be losing,
  we have $a(2k+3)+b(k+1)=t-1$. As mentioned before, the unique solution of this equation is given by $a=k$ and $b=0$.
  Thus, we have $\Shift_1(v)=1$ for $k=1$ and $\Shift_1(v)=2$ for $k>1$.

  Every minimal winning coalition besides $\{1,2\}$ containing {\voter}~$2$ has to be of type $(0,1,a,b)$. For
  any $0\le a\le k$ the unique value for $b$ is given by $b(a)=\left\lceil\frac{t-a(2k+3)}{k+1}\right\rceil$, where
  $1\le b\le 2k+1$. Since $2k+3>2\cdot(k+1)$ we have $b(a)+1<b(a-1)$, i.e., the corresponding coalitions are shift-minimal
  winning. Counting the number of coalitions of type $(0,1,a,b(a))$ gives
  $$
    \Shift_2(v)=-1+\Shift_1(v)+\sum_{a=0}^{k}\sum_{b=\left\lceil\frac{t-a(2k+3)}{k+1}\right\rceil}
    {k \choose a}\cdot{{2k+1+m}\choose b}.
  $$
\end{Proof}
Of course one may speculate whether the stated construction is tight in general. At the very least we can conclude
that $c_{\mathcal{P}}(n,\mathfrak{W})$ tends at least exponentially to $1$, i.e., there are constants $d_1>0$, $d_2>1$
with $c_{\mathcal{P}}(n,\mathfrak{W})\ge 1-d_1\cdot d_2^{-n}$, for $\mathcal{P}=\left(\BZ,\Shift\right)$.
Conjecture~\ref{conj_BZ_PGI_weighted} on the other hand would imply only a linear convergence rate. So, in some sense, the
Shift index is even less locally monotonic than the Public Good index.

Next we go on and consider restrictions of the class of weighted games.

\begin{lemma}
  \label{lemma_PGI_proper}
  For $\mathcal{P}=\left(\BZ,\PGI\right)$ and $n\ge 3$ we have
  $c_{\mathcal{P}}(n,\mathfrak{W}^p)\ge \max\!\left(0,\frac{n-4}{n-2}\right)$.
\end{lemma}
\begin{Proof}
  Since $c_{\mathcal{P}}(n,\mathfrak{W})\ge 0$ by definition, it suffices to consider weighted games with $n\ge 5$ {\voter}s.
  We consider the weighted game $v=[2n-3;n-1,n-2,n-2,1,\dots,1]$ with $n-3$ {\voter}s of weight $1$, two {\voter}s of
  weight $n-2>1$ and one {\voter} of weight $n-1$. Since the sum of voting weights is $4n-8<2\cdot (2n-3)$, the game
  is proper. The minimal winning coalitions are given by $\{1,2\}$, $\{1,3\}$,
  and $\{2,3,i\}$ for all $4\le i\le n$. Thus, we have $\PGI(v)=(2,n-2,n-2,1,\dots,1)$. The swing coalitions for {\voter}~$1$
  are given by $\{1,2,3\}$ and all coalitions of the form $\{1,i\}\cup S$, where $i\in\{2,3\}$ and $S\subseteq \{4,\dots,n\}$.
  The swing coalitions for {\voter}~$2$ are given by $\{1,2\}$ and all coalitions of the form $\{2,3\}\cup S$, where
  $\emptyset\neq S\subseteq\{4,\dots,n\}$. The unique swing coalition for a {\voter} $i\ge 4$ is given by  $\{2,3,i\}$.
  Thus, we have $\BZ(v)=\left(2^{n-2}+1,2^{n-2}-1,2^{n-2}-1,1,\dots,1\right)$.

  For {\voter}s~$1$, $2$ and game $v$ Inequality~(\ref{ie_feasible}) is equivalent to
  $2\alpha_1 \ge \alpha_1\cdot \BZ_2(v)+(n-4)\alpha_2$. Inserting $\alpha_1+\alpha_2=1$ yields $\alpha_1\ge \frac{n-4}{n-2}$.
\end{Proof}

\begin{corollary}
  \label{cor_PGI_proper}
  For $\mathcal{P}=\left(\BZ,\Shift\right)$ and $n\ge 3$ we have
  $c_{\mathcal{P}}(n,\mathfrak{W}^p)\ge \max\!\left(0,\frac{n-4}{n-2}\right)$.
\end{corollary}

Using the ILP approach we can verify that Lemma~\ref{lemma_PGI_proper} is tight for $n\le 10$ and
Corollary~\ref{cor_PGI_proper} is tight for $n\le 6$.

\begin{lemma}
  \label{lemma_BZ_Shift_proper_special}
  Let $\mathcal{P}=\left(\BZ,\Shift\right)$.
  \begin{enumerate}
    \item[(1)] For $n=7$ we have $c_{\mathcal{P}}(n,\mathfrak{W}^p)= \frac{5}{7}   \approx 0.71429$.
    \item[(2)] For $n=8$ we have $c_{\mathcal{P}}(n,\mathfrak{W}^p)= \frac{11}{13} \approx 0.84615$.
    \item[(3)] For $n=9$ we have $c_{\mathcal{P}}(n,\mathfrak{W}^p)= \frac{12}{13} \approx 0.92308$.
    \item[(4)] For $n=10$ we have $c_{\mathcal{P}}(n,\mathfrak{W}^p)= \frac{21}{22}\approx 0.95455$.
    \item[(5)] For $n=11$ we have $c_{\mathcal{P}}(n,\mathfrak{W}^p)= \frac{83}{85}\approx 0.97647$.
  \end{enumerate}
\end{lemma}
\begin{Proof}
  For the lower bounds we state an explicit weighted game and the Banzhaf and the Shift score for the first
  two {\voter}s:
  \begin{enumerate}
    \item [(1)]$[21;11,10,5,5,3,3,3]$, $\BZ=(33,31,\dots)$, $\Shift=(2,7,\dots)$;
    \item [(2)]$[25;13,12,5,5,3,3,3,3]$, $\BZ=(65,63,\dots)$, $\Shift=(2,13,\dots)$;
    \item [(3)]$[31;16,15,7,7,3,3,3,3,3]$, $\BZ=(129,127,\dots)$, $\Shift=(2,26,\dots)$;
    \item [(4)]$[39;12,11,9,9,9,5,5,5,5,5]$, $\BZ=(194,192,\dots)$, $\Shift=(1,45,\dots)$;
    \item [(5)]$[32;8,7,7,7,7,4,4,4,4,4,4]$, $\BZ=(324,322,\dots)$, $\Shift=(1,84,\dots)$.
  \end{enumerate}
  For the upper bounds we have applied the ILP approach.
\end{Proof}

Since the parametric example from Lemma~\ref{lemma_lower_bound_PGI_weighted} and the examples
from Lemma~\ref{lemma_lower_bound_Shift_weighted_special} and Lemma~\ref{lemma_general_construction} have
the property that the sum of weights
meets or exceeds twice the quota, the respective games are all strong. Thus, the same results
are valid if we restrict the class of weighted games to weighted strong games. Clearly we also conjecture
$c_{\mathcal{P}}(n,\mathfrak{W}^s)=\max\!\left(0,\frac{n-3}{n-1}\right)$ for all $n\ge 2$, where
$\mathcal{P}=\left(\BZ,\PGI\right)$, which is a weakening of Conjecture~\ref{conj_BZ_PGI_weighted}.

Since the parametric example from Lemma~\ref{lemma_PGI_proper} and the examples from Lemma~\ref{lemma_BZ_Shift_proper_special}
are not strong, the cost of local monotonicity may decrease for constant sum weighted games compared to proper weighted games.

If the class of weighted games is restricted to constant-sum games, then the non-monotonicity is generally reduced, i.e.\
a lower contribution of the Banzhaf score is sufficient to guarantee monotonicity of the power index obtained as a
convex combination.

\begin{lemma}
  \label{lemma_lower_bound_PGI_constant_sum}
  For $\mathcal{P}=\left(\BZ,\PGI\right)$ and $n\ge 2$ we have
  $c_{\mathcal{P}}(n,\mathfrak{W}^c)\ge \max\!\left(0,\frac{n-5}{n-1}\right)$.
\end{lemma}
\begin{Proof}
  Since $c_{\mathcal{P}}(n,\mathfrak{W})\ge 0$ by definition, it suffices to consider weighted games with $n\ge 6$ {\voter}s.
  For the weighted game $v=[2n-5;n-2,n-3,n-3,1\dots,1]$, with $n-3$ players of weight $1$, the minimal winning
  coalitions are given by $\{1,2\}$, $\{1,3\}$, $\{1,4,5,\dots,n\}$, and $\{2,3,i\}$ , where $4\le i\le n$. Thus,
  we have $\operatorname{PGI^S}_1(v)=3$ and $\operatorname{PGI^S}_2(v)=n-2$. For player~$1$ the swing coalitions
  are given by $\{1,2,3\}$, $\{1,4,5,\dots,n\}$, and $\{1,2\}\cup S$, $\{1,3\}\cup S$, where $S\subseteq \{4,5,\dots,n\}$.
  For player~$2$ the swing coalitions are given by $\{1,2\}\cup U$, where $U \subseteq \{4,5,\dots,n\}$ with $|U|<n-3$, and
  $\{2,3\}\cup V$, where $\emptyset \neq V \subseteq \{4,5,\dots,n\}$. Thus, we have $\operatorname{BZ^S}_1(v)=2^{n-2}+2$
  and $\operatorname{BZ^S}_2(v)=2^{n-2}-2$.

  For players~$1$, $2$ and game $v$ Inequality~(\ref{ie_feasible}) is equivalent to
  $4\alpha_1 \ge \alpha_1\cdot \BZ_2(v)+(n-5)\alpha_2$. Inserting $\alpha_1+\alpha_2=1$ yields $\alpha_1\ge \frac{n-5}{n-1}$.
\end{Proof}

\begin{corollary}
  \label{cor_lower_bound_Shift_constant_sum}
  For $\mathcal{P}=\left(\BZ,\Shift\right)$ and $n\ge 2$ we have
  $c_{\mathcal{P}}(n,\mathfrak{W}^c)\ge \max\!\left(0,\frac{n-5}{n-1}\right)$.
\end{corollary}

\begin{corollary}
  $$
    \lim_{n\to\infty} c_{\left(\BZ,\PGI\right)}(n,\mathfrak{W}^c)=
    \lim_{n\to\infty} c_{\left(\BZ,\Shift\right)}(n,\mathfrak{W}^c)=
    \lim_{n\to\infty} c_{\left(\BZ,\PGI,\Shift\right)}(n,\mathfrak{W}^c)=1
  $$
\end{corollary}

Using the ILP approach we can verify that Lemma~\ref{lemma_lower_bound_PGI_constant_sum} is tight for $n\le 11$ and
Corollary~\ref{cor_lower_bound_Shift_constant_sum} is tight for $n\le 7$.

\begin{lemma}
  Let $\mathcal{P}=\left(\BZ,\Shift\right)$.
  \begin{enumerate}
    \item[(1)] For $n=8$ we have $c_{\mathcal{P}}(n,\mathfrak{W}^c)= \frac{2}{3}   \approx 0.66667$.
    \item[(2)] For $n=9$ we have $c_{\mathcal{P}}(n,\mathfrak{W}^c)= \frac{23}{27} \approx 0.85185$.
    \item[(3)] For $n=10$ we have $c_{\mathcal{P}}(n,\mathfrak{W}^c)= \frac{43}{47}\approx 0.91489$.
    \item[(4)] For $n=11$ we have $c_{\mathcal{P}}(n,\mathfrak{W}^c)= \frac{75}{79}\approx 0.94937$.
  \end{enumerate}
\end{lemma}
\begin{Proof}
  For the lower bounds we state an explicit weighted game and the Banzhaf and the Shift score for the first
  two {\voter}s:
  \begin{enumerate}
    \item [(1)]$[17;9,8,5,3,2,2,2,2]$, $\BZ=(66,62,\dots)$, $\Shift=(3,11,\dots)$;
    \item [(2)]$[21;11,10,5,5,2,2,2,2,2]$, $\BZ=(130,126,\dots)$, $\Shift=(3,26,\dots)$;
    \item [(3)]$[21;6,5,5,5,5,3,3,3,3,3]$, $\BZ=(170,166,\dots)$, $\Shift=(5,48,\dots)$;
    \item [(4)]$[22;8,7,7,7,2,2,2,2,2,2,2]$, $\BZ=(386,382,\dots)$, $\Shift=(4,79,\dots)$.
  \end{enumerate}
  For the upper bounds we have applied the ILP approach.
\end{Proof}

Similar results can be obtained for $\J$, $\DP$, and $\SDP$.

\begin{lemma}
  \label{lemma_J_DP}
  Let $\mathcal{P}=\left(\J,\DP\right)$.
  \begin{enumerate}
    \item[(1)] For $n\le 4$ we have $c_{\mathcal{P}}(n,\mathfrak{W})= 0$.
    \item[(2)] For $n=5$ we have $c_{\mathcal{P}}(n,\mathfrak{W})= \frac{1}{8}=0.125$.
    \item[(3)] For $n=6$ we have $c_{\mathcal{P}}(n,\mathfrak{W})= \frac{1}{4}=0.25$.
    \item[(4)] For $n=7$ we have $c_{\mathcal{P}}(n,\mathfrak{W})= \frac{1}{3}\approx 0.33333$.
    \item[(5)] For $n=8$ we have $c_{\mathcal{P}}(n,\mathfrak{W})= \frac{2}{5}=0.4$.
    \item[(6)] For $n=9$ we have $c_{\mathcal{P}}(n,\mathfrak{W})\ge\frac{11}{25}=0.44$.
    \item[(7)] For $n=10$ we have $c_{\mathcal{P}}(n,\mathfrak{W})\ge\frac{1}{2}=0.5$.
  \end{enumerate}
\end{lemma}
\begin{Proof}
  For the lower bounds we state an explicit weighted game, the Johnston and the Deegan-Packel score, where we highlight
  the values of the critical {\voter}s:
  \begin{enumerate}
    \item [(2)] $[3;3,2,2,1,1]$, $\J=(\mathbf{6},\mathbf{\frac{5}{2}},\frac{5}{2},1,1)$, $\SDP=(\mathbf{1},\mathbf{\frac{3}{2}},\frac{3}{2},1,1)$;
    \item [(3)] $[8;4,4,3,1,1,1]$, $\J=(\frac{15}{2},\mathbf{\frac{15}{2}},\mathbf{6},\frac{2}{3},\frac{2}{3},\frac{2}{3})$,
                $\SDP=(\frac{3}{2},\mathbf{\frac{3}{2}},\mathbf{2},\frac{2}{3},\frac{2}{3},\frac{2}{3})$;
    \item [(4)] $[9;5,4,3,2,2,2,2]$, $\J=(\frac{70}{3},\mathbf{\frac{28}{3}},\mathbf{\frac{23}{3}},\frac{19}{6},\frac{19}{6},\frac{19}{6},\frac{19}{6})$,\\
                $\SDP=(\frac{23}{6},\mathbf{\frac{17}{6}},\mathbf{\frac{11}{3}},\frac{19}{6},\frac{19}{6},\frac{19}{6},\frac{19}{6})$;
    \item [(5)] $[12;4,4,4,3,3,3,3,3]$, $\J=(\frac{19}{2},\frac{19}{2},\mathbf{\frac{19}{2}},\mathbf{\frac{17}{2}},\frac{17}{2},\frac{17}{2},\frac{17}{2},\frac{17}{2})$,\\
                $\SDP=(\frac{47}{6},\frac{47}{6},\mathbf{\frac{47}{6}},\mathbf{\frac{17}{2}},\frac{17}{2},\frac{17}{2},\frac{17}{2},\frac{17}{2})$;
    \item [(6)] $[20;5,5,5,5,4,4,4,4,4]$, $\J=(\frac{19}{2},\dots,\mathbf{\frac{19}{2}},\mathbf{\frac{69}{5}},\dots,\frac{69}{5})$,\\
                $\SDP=(\frac{53}{4},\dots,\mathbf{\frac{53}{4}},\mathbf{\frac{69}{5}},\dots,\frac{69}{5})$;
    \item [(7)] $[20;5,5,5,5,4,4,4,4,4,4]$, $\J=(\frac{103}{4},\dots,\mathbf{\frac{103}{4}},\mathbf{25},\dots,25)$,\\
                $\SDP=(\frac{97}{4},\dots,\mathbf{\frac{97}{4}},\mathbf{25},\dots,25)$;
  \end{enumerate}
  For the upper bounds we have applied the ILP approach.
\end{Proof}

For $\mathcal{P}=\left(\J,\DP\right)$ the cost of local monotonicity seems to be increasing rather slowly. Given the numerical
data from Lemma~\ref{lemma_J_DP} it is not clear at all whether $c_{\mathcal{P}}(n,\mathfrak{W})$ tends to $1$ as $n$ tends to
infinity. To this end we consider the following construction for a odd number of {\voter}s:

\begin{lemma}
  \label{construction_J_DP}
  For $k\ge 1$ and $v=[k(k+1),\overset{k\text{ times}}{\overbrace{k+1,\dots,k+1}},\overset{k+1\text{ times}}{\overbrace{k,\dots,k}}]$
  we have $\J_i(v)=c(k)$ for all $1\le i\le k$, $\J_i(v)=d(k)$ for all $k+1\le 2k+1$, $\DP_i(v)=c(k)-\frac{k+1}{k}$ for all $1\le i\le k$,
  and $\DP_i(v)=d(k)$ for all $k+1\le 2k+1$, where
  $$
    c(k)=\frac{k+2}{k}+\frac{1}{k+1}\sum_{i=1}^{k-1} {{k-1}\choose{i-1}}\cdot {{k+1}\choose{i}}
  $$
  and
  $$
    d(k)=\frac{1}{k+1}+\frac{1}{k+1}\cdot\sum_{i=1}^{k-1} {k\choose{i}}^2.
  $$
\end{lemma}
\begin{Proof}
  We can easily check that $v$ consists of $n=2k+1$ {\voter}s having two different weights. As in the proof of
  Lemma~\ref{lemma_general_construction} we use a $2$-tuple $(c_1,c_2)$ to describe the type of a coalition.
  Since the coalitions of type $(k,0)$ are winning but the coalitions of type $(0,k)$ are losing, no {\voter}
  of weight $k+1$ is equivalent to a player of weight $k$.

  In coalitions of type $(k,0)$ or type $(k,1)$ all {\voter}s of weight $k+1$ are swing {\voter}s, while
  the {\voter}s of weight $k$ are not swing {\voter}s. The other types of coalitions which contain a least
  one swing {\voter} are given by $(i,k+1-i)$, where $0\le i\le k-1$. In these cases all involved $k+1$ {\voter}s
  are swings.
  Counting the number of cases, where {\voter} $1$ is contained, for each of the mentioned coalition types
  gives
  \begin{eqnarray*}
    \J_1(v)&=&{{k-1}\choose{k-1}}\cdot {{k+1}\choose{0}}\cdot\frac{1}{k}+
            {{k-1}\choose{k-1}}\cdot {{k+1}\choose{1}}\cdot\frac{1}{k}\\
           &&+\frac{1}{k+1}\cdot\sum_{i=1}^{k-1} {{k-1}\choose{i-1}}\cdot{{k+1}\choose{i}}\\
           &=&c(k)
  \end{eqnarray*}
  Counting the number of cases, where {\voter} $n$ is contained, for each of the mentioned coalition types
  gives
  \begin{eqnarray*}
    \J_{n}(v)&=& \frac{1}{k+1}+\frac{1}{k+1}\cdot\sum_{i=1}^{k-1} {k\choose{i}}\cdot{{k}\choose{i}}=d(k).
  \end{eqnarray*}
  All coalition types except $(k,1)$ correspond to minimal winning coalitions. Thus we have
  $\DP_1(v)=\J_1(v)-{{k-1}\choose{k-1}}\cdot {{k+1}\choose{1}}\cdot\frac{1}{k}=c(k)-\frac{k+1}{k}$ and
  $\DP_n(v)=\J_n(v)=d(k)$. The values for the remaining {\voter}s follow from symmetry.
\end{Proof}

\begin{corollary}
  For $\mathcal{P}=\left(\J,\DP\right)$ and $n\ge 1$ we have $c_{\mathcal{P}}(n,\mathfrak{W})\ge 1-\frac{2(3n-2)}{n^2}\ge 1-\frac{6}{n}$.
\end{corollary}
\begin{Proof}
  For the weighted game from Lemma~\ref{construction_J_DP} Inequality~(\ref{ie_feasible}) yields
  $c_{\mathcal{P}}(2k+1,\mathfrak{W})\ge 1-\frac{k}{k+1}\cdot(c(k)-d(k))$. Since
  $\sum_{i=1}^{k-1} {{k-1}\choose{i-1}}\cdot {{k+1}\choose{i}}-\sum_{i=1}^{k-1} {k\choose{i}}^2=-(k-1)$,
  we have $c(k)-d(k)=\frac{3k+2}{k(k+1)}$. Thus $c_{\mathcal{P}}(2k+1,\mathfrak{W})\ge 1-\frac{3k+2}{(k+1)^2}$.
  Since $c_{\mathcal{P}}(n,\mathfrak{W})\ge c_{\mathcal{P}}(n-1,\mathfrak{W})$ we can choose $k=\left\lceil\frac{n-2}{2}\right\rceil$
  and obtain the stated lower bounds.
\end{Proof}

\begin{corollary}
  For $\mathcal{P}=\left(\J,\DP\right)$ we have $\lim\limits_{n\to\infty}c_{\mathcal{P}}(n,\mathfrak{W})=1$.
\end{corollary}

\begin{lemma}
  Let $\mathcal{P}=\left(\J,\SDP\right)$.
  \begin{enumerate}
    \item[(1)] For $n\le 4$ we have $c_{\mathcal{P}}(n,\mathfrak{W})= 0$.
    \item[(2)] For $n=5$ we have $c_{\mathcal{P}}(n,\mathfrak{W})= \frac{1}{3}    \approx 0.33333$.
    \item[(3)] For $n=6$ we have $c_{\mathcal{P}}(n,\mathfrak{W})= \frac{3}{5}=0.6$.
    \item[(4)] For $n=7$ we have $c_{\mathcal{P}}(n,\mathfrak{W})= \frac{7}{9}    \approx 0.77778$.
    \item[(5)] For $n=8$ we have $c_{\mathcal{P}}(n,\mathfrak{W})= \frac{47}{53}\approx 0.88679$.
    \item[(6)] For $n=9$ we have $c_{\mathcal{P}}(n,\mathfrak{W})\ge \frac{29}{31}\approx 0.93548$.
  \end{enumerate}
\end{lemma}
\begin{Proof}
  For the lower bounds we state an explicit weighted game and the Johnston and the Shift Deegan-Packel score for the last
  two {\voter}s:
  \begin{enumerate}
    \item [(2)]$[4;3,3,2,2,1]$, $\J=(\dots,2,1)$, $\SDP=(\dots,\frac{1}{2},1)$;
    \item [(3)]$[8;4,4,3,3,2,1]$, $\J=(\dots,2,\frac{4}{3})$, $\SDP=(\dots,\frac{1}{3},\frac{4}{3})$;
    \item [(4)]$[8;5,5,2,2,2,2,1]$, $\J=(\dots,\frac{19}{6},\frac{8}{3})$, $\Shift=(\dots,\frac{11}{12},\frac{8}{3})$;
    \item [(5)]$[15;7,7,3,3,3,3,3,2]$, $\J=(\dots,\frac{86}{15},\frac{16}{3})$, $\SDP=(\dots,\frac{11}{5},\frac{16}{3})$;
    \item [(6)]$[12;7,7,2,2,2,2,2,2,1]$, $\J=(\dots,\frac{47}{6},\frac{15}{2})$, $\SDP=(\dots,\frac{8}{3},\frac{15}{2})$.
  \end{enumerate}
  For the upper bounds we have applied the ILP approach.
\end{Proof}

Quite obviously the cost of local monotonicity for $\mathcal{P}=\left(\J,\DP\right)$ seems to converge to $1$ as $n$ increases.
An appropriate lower bound can be concluded from the parametric example
$v=[2(n-3);n-2,n-2,\underset{n-3\text{ times}}{\underbrace{2,\dots,2}},1]$, where $n\ge 5$, by considering the last two
{\voter}s. We remark that the exact value for $n=7$ and the lower bound for $n=9$ is attained for this parametric family.

\section{Determining the polyhedron $\mathbb{P}^{\mathcal{P}}_{\text{LM}}(n,\mathfrak{W})$ for convex combinations
of three power indices.}
\label{sec_polyhedron}

\noindent
In the previous section we have computationally determined the cost of local monotonicity for several sets of two or three power
indices on subclasses of weighted games. Now we want to gain even more information: Given a collection $\mathcal{P}$
of $r\ge 2$ power indices, for which $\alpha\in \mathbb{S}^r$ does $P^{\alpha,\mathcal{P}}$ satisfy local
monotonicity? In Section~\ref{sec_convex_combinations} we have obtained the result that the respective set
$\mathbb{P}^{\mathcal{P}}_{\text{LM}}(n,\mathfrak{G})$ is a polyhedron. As already discussed, each game $v\in \mathfrak{G}$
gives a valid inequality for $\mathbb{P}^{\mathcal{P}}_{\text{LM}}(n,\mathfrak{G})$. Using the ILP approach from
Section~\ref{sec_ILP} we can check whether a given point $\alpha\in\mathbb{S}^r$ is contained in
$\mathbb{P}^{\mathcal{P}}_{\text{LM}}(n,\mathfrak{G})$. In the case where $\alpha$ is not contained in
$\mathbb{P}^{\mathcal{P}}_{\text{LM}}(n,\mathfrak{G})$, we obtain a game $v\in\mathfrak{G}$ verifying this fact. So either
we can verify vertices of our polyhedron or compute additional non-redundant valid inequalities. So, instead of looping over
all games in $\mathfrak{G}$, we can use the following algorithm to determine $\mathbb{P}^{\mathcal{P}}_{\text{LM}}(n,\mathfrak{G})$:

\bigskip

\noindent
$\mathbb{P}=\mathbb{S}^r$\\
compute the set $\mathcal{A}$ of vertices of $\mathbb{P}$\\
\textbf{for all} $\alpha\in\mathcal{A}$ \textbf{do}\\
\hspace*{5mm} $verified(\alpha)=false$\\
\textbf{end for}
\textbf{while} $\exists \alpha\in \mathcal{A}$ with $verified(\alpha)=false$ \textbf{do}\\
\hspace*{5mm} \textbf{if} $\alpha\in \mathbb{P}^{\mathcal{P}}_{\text{LM}}(n,\mathfrak{G})$ \textbf{then}\\
\hspace*{10mm} $verified(\alpha)=true$\\
\hspace*{5mm} \textbf{else}\\
\hspace*{10mm} compute certifying game $v\in\mathfrak{G}$\\
\hspace*{10mm} add inequalities corresponding to $v$ to $\mathbb{P}$\\
\hspace*{10mm} compute the set $\mathcal{A}$ of vertices of $\mathbb{P}$\\
\hspace*{10mm} set $verified(\alpha)=false$ for all new vertices\\
\hspace*{5mm}\textbf{end if}\\
\textbf{end while}\\
\textbf{return} $\mathbb{P}$

\medskip

We remark that we may also include the information that $(1,0,\dots,0)\in \mathbb{P}^{\mathcal{P}}_{\text{LM}}(n,\mathfrak{G})$
in any case, i.e., one of the $r$ vertices of $\mathbb{S}^r$ can be set to be verified. If already determined, the $r-1$ examples
for the cost of local monotonicity for $\mathcal{P}'=\left\{P^1,P^i\right\}$ can be used to replace the initialization
of $\mathbb{P}$, i.e., setting $\mathbb{P}=\operatorname{conv}(e_1,p_2,\dots,p_r)$, where $e_i$ is the $i$th unit vector
and $p_i=e_1\cdot c_{\left\{P^1,P^i\right\}}(n,\mathfrak{G})+e_i\cdot (1-c_{\left\{P^1,P^i\right\}}(n,\mathfrak{G}))$.

Exemplarily, we have performed the computations for $\mathfrak{G}=\mathfrak{W}$, $\mathcal{P}=\{\BZ,\PGI,\Shift\}$, and
$n\le 9$.

\begin{lemma}
  For $\mathcal{P}=\{\BZ,\PGI,\Shift\}$ we have
  \begin{itemize}
    \item[(1)] $\mathbb{P}^{\mathcal{P}}_{\text{LM}}(n,\mathfrak{W})=\mathbb{S}^3=\operatorname{conv}\left\{(1,0,0),(0,1,0),(0,0,1)\right\}$
               for $n\le 3$;
    \item[(2)] $\mathbb{P}^{\mathcal{P}}_{\text{LM}}(4,\mathfrak{W})=\operatorname{conv}\left\{(1,0,0),(\frac{1}{3},\frac{2}{3},0),(\frac{1}{3},0,\frac{2}{3})\right\}$;
    \item[(3)] $\mathbb{P}^{\mathcal{P}}_{\text{LM}}(5,\mathfrak{W})=\operatorname{conv}\left\{(1,0,0),(\frac{1}{2},\frac{1}{2},0),(\frac{1}{2},0,\frac{1}{2})\right\}$;
    \item[(4)] $\mathbb{P}^{\mathcal{P}}_{\text{LM}}(6,\mathfrak{W})=\operatorname{conv}\left\{(1,0,0),(\frac{3}{5},\frac{2}{5},0),(\frac{3}{5},0,\frac{2}{5})\right\}$;
    \item[(5)] $\mathbb{P}^{\mathcal{P}}_{\text{LM}}(7,\mathfrak{W})=\operatorname{conv}\left\{(1,0,0),(\frac{2}{3},\frac{1}{3},0),(\frac{7}{9},0,\frac{2}{9}),(\frac{2}{3},\frac{1}{4},\frac{1}{12})\right\}$;
    \item[(6)] $\mathbb{P}^{\mathcal{P}}_{\text{LM}}(8,\mathfrak{W})=\operatorname{conv}\left\{(1,0,0),(\frac{5}{7},\frac{2}{7},0),(\frac{7}{8},0,\frac{1}{8}),(\frac{5}{7},\frac{9}{35},\frac{1}{35})\right\}$;
    \item[(7)] $\mathbb{P}^{\mathcal{P}}_{\text{LM}}(9,\mathfrak{W})=\operatorname{conv}\left\{(1,0,0),(\frac{3}{4},\frac{1}{4},0),(\frac{25}{27},0,\frac{2}{27}),(\frac{3}{4},\frac{19}{80},\frac{1}{80})\right\}$.
  \end{itemize}
\end{lemma}

\begin{figure}[htp]
\begin{center}
  \includegraphics{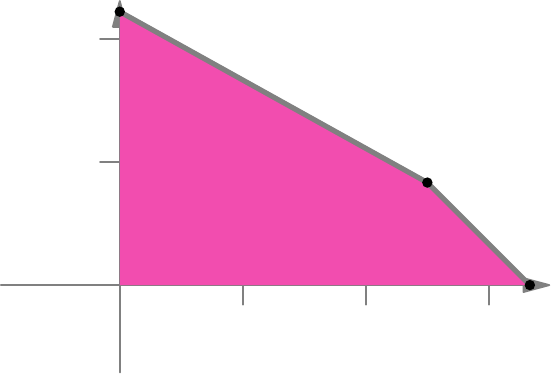}
\caption{$\mathbb{P}^{\mathcal{P}}_{\text{LM}}(7,\mathfrak{W})$ for $\mathcal{P}=\{\BZ,\PGI,\Shift\}$.}
\label{fig_compatible_n_7}
\end{center}
\end{figure}

In Figure~\ref{fig_compatible_n_7} we have exemplarily drawn $\mathbb{P}^{\{\BZ,\PGI,\Shift\}}_{\text{LM}}(7,\mathfrak{W})$, which
complements the region drawn in Figure~\ref{fig_non_compatible_n_7}. In order to illustrate the proposed algorithm we consider the
case $n=9$ as an example. For $\{\BZ,\PGI\}$ the cost of local monotonicity is given by $\frac{3}{4}$ and e.g.\ attained at the
game $v_1=[2;2,1,1,1,1,1,1,1,1]$. We have $\BZ(v_1)=(9,7,\dots)$, $\PGI(v_1)=(1,7,\dots)$, and $\Shift(v_1)=(1,7,\dots)$.
For $\{\BZ,\Shift\}$ the cost of local monotonicity is given by $\frac{25}{27}$ and e.g.\ attained at the
game $v_2=[30;16,15,7,7,3,3,3,3,3]$. We have $\BZ(v_2)=(129,127,\dots)$, $\PGI(v_2)=(23,27,\dots)$, and $\Shift(v_2)=(2,27,\dots)$.
The hyperplane corresponding to $v_1$ is given by $2\alpha_1-6\alpha_2-6\alpha_3=0$ and the hyperplane corresponding to
$v_2$ is given by $2\alpha_1-4\alpha_2-25\alpha_3$. Together with $\alpha_1+\alpha_2+\alpha_3=1$ we obtain the new vertex
$\alpha'=\left(\frac{3}{4},\frac{19}{84},\frac{1}{42}\right)$. By using the ILP approach we can compute that $\alpha'$ does
not lead to a locally monotonic power index and obtain the game $v_3=[18;13,12,5,2,2,2,2,2,2]$ with $\BZ(v_3)=(129,127,\dots)$,
$\PGI(v_3)=(22,27,\dots)$, and $\Shift(v_3)=(1,26,\dots)$. For this game the corresponding hyperplane is given by
$2\alpha_1-5\alpha_2-25\alpha_3$. Again there arises exactly one new vertex -- $\alpha''=\left(\frac{3}{4},\frac{19}{80},\frac{1}{80}\right)$.
By using the ILP approach we can compute that $\alpha''$ is contained in $\mathbb{P}^{\mathcal{P}}_{\text{LM}}(9,\mathfrak{W})$, so
that the determination of the polyhedron is completed.

We remark that $\alpha''$ does also attain the cost of local monotonicity for $\{\BZ,\Shift\}$. So if we had started with
the games $v_1$ and $v_3$ instead of $v_1$ and $v_2$, our algorithm would have needed one iteration less.

\section{Conclusion}
\label{sec_conclusion}

\noindent
We have introduced the concept of considering convex combinations of power indices. Several of
the main properties of power indices are preserved by convexity, i.e., given a collection $\mathcal{P}$
of power indices such that each power index in $\mathcal{P}$ has a certain property, then also every
convex combination of the power indices in $\mathcal{P}$ has this property. Lemma~\ref{lemma_properties}
gives some examples of such properties being preserved by convexity. The freedom in choosing the multipliers
almost arbitrarily enables us to search for power indices, which satisfy some other useful properties.
As an application we study local monotonicity. It is well known that the Public Good index does not
satisfy local monotonicity, while e.g.\ the Banzhaf index does. So, what proportion of the Banzhaf index
is necessary so that a convex combination of both indices becomes locally monotonic? The newly introduced
cost of local monotonicity answers this specific question. Similar measures may of course be introduced
for other properties of power indices.

It turns out that with an increasing number of {\voter}s the weight of the Banzhaf index needs to tend to $1$.
For a finite number of {\voter}s there is still some freedom to incorporate some information from
the Public Good index, while maintaining the local monotonicity of the Banzhaf index.

Restricting the class of the underlying games to strong or proper games typically decreases the cost of local
monotonicity, but does not change the general behavior.

The cost of local monotonicity for combinations of the Banzhaf and the Shift index is considerably higher. So, in some
sense the Shift index is even less locally monotonic than the Public Good index. Going over to the so-called \textit{equal division}
version of the Banzhaf-, Public Good-, and the Shift index, i.e., The Johnston-, the Deegan-Packel and the Shift-Deegan Packel
index, seems to generally lower the cost of local monotonicity. Nevertheless, the corresponding cost of local monotonicity
approaches $1$ as the number of {\voter}s tends to infinity.

So, our study has shed some light on the property of local monotonicity of some power indices. The methodology of
considering convex combinations of power indices in order to obtain some desirable properties is quite general
and the presented theoretical and algorithmic framework may be applied in further studies.



\begin{thebibliography}{10}

\bibitem{EU_power_distribution}
E.~Algaba, J.M. Bilbao, and J.R. Fern\'andez.
\newblock The distribution of power in the {E}uropean {C}onstitution.
\newblock {\em European J. Oper. Res.}, 176(3):1752--1766, 2007.


\bibitem{alonso2010new}
J.M. Alonso-Meijide and J.~Freixas.
\newblock A new power index based on minimal winning coalitions without any
  surplus.
\newblock {\em Decision Support Systems}, 49(1):70--76, 2010.

\bibitem{AlonsoMeijide20123395}
J.M. Alonso-Meijide, J.~Freixas, and X.~Molinero.
\newblock Computation of several power indices by generating functions.
\newblock {\em Appl. Math. Comput.}, 219(8):3395--3402, 2012.

\bibitem{banzhaf1964weighted}
J.F. Banzhaf.
\newblock Weighted voting doesn't work: {A} mathematical analysis.
\newblock {\em Rutgers Law Rev.}, 19:317--343, 1965.

\bibitem{bertini2013comparing}
C.~Bertini, J.~Freixas, G.~Gambarelli, and I.~Stach.
\newblock Comparing power indices.
\newblock {\em Int. Game Theory Rev.}, 15(2), 2013.

\bibitem{EU_enlargement}
J.M. Bilbao, J.R. Fern\'andez, N.~Jim\'enez, and J.J. L\'opez.
\newblock Voting power in the {E}uropean {U}nion enlargement.
\newblock {\em European J. Oper. Res.}, 143(1):181--196, 2002.

\bibitem{deegan1978new}
J.~Deegan~Jr and E.W. Packel.
\newblock A new index of power for simple $n$-person games.
\newblock {\em Internat. J. Game Theory}, 7(2):113--123, 1978.

\bibitem{felsenthal1995postulates}
D.S. Felsenthal and M.~Machover.
\newblock Postulates and paradoxes of relative voting power -- a critical
  re-appraisal.
\newblock {\em Theory and Decision}, 38(2):195--229, 1995.

\bibitem{0954.91019}
D.S. Felsenthal and M.~Machover.
\newblock {\em The measurement of voting power: {T}heory and practice, problems
  and paradoxes}.
\newblock Cheltenham: Edward Elgar. xviii, 322 p., 1998.

\bibitem{freixas1997common}
J.~Freixas and G.~Gambarelli.
\newblock Common internal properties among power indices.
\newblock {\em Control and Cybernetics}, 26(4):591--603, 1997.

\bibitem{freixas2011alpha}
J.~Freixas and S.~Kurz.
\newblock On $\alpha$-roughly weighted games.
\newblock {\em Internat. J. Game Theory}, 43(3):659--692, 2014.

\bibitem{freixas2012ordinal}
J.~Freixas, D.~Marciniak, and M.~Pons.
\newblock On the ordinal equivalence of the {J}ohnston, {B}anzhaf and {S}hapley
  power indices.
\newblock {\em European J. Oper. Res.}, 216(2):367--375,
  2012.

\bibitem{freixas2010without}
J.~Freixas and X.~Molinero.
\newblock Weighted games without a unique minimal representation in integers.
\newblock {\em Optim. Methods Softw.}, 25:203--215, 2010.

\bibitem{1151.91021}
J.~Freixas and M.~A. Puente.
\newblock Dimension of complete simple games with minimum.
\newblock {\em European J. Oper. Res.}, 188(2):555--568, 2008.

\bibitem{holler1982forming}
M.J. Holler.
\newblock Forming coalitions and measuring voting power.
\newblock {\em Political studies}, 30(2):262--271, 1982.

\bibitem{holler2004monotonicity}
M.J. Holler and S.~Napel.
\newblock Monotonicity of power and power measures.
\newblock {\em Theory and Decision}, 56(1-2):93--111, 2004.

\bibitem{holler2013reflections}
M.J. Holler and H.~Nurmi.
\newblock Reflections on power, voting, and voting power.
\newblock In M.J. Holler and H.~Nurmi, editors, {\em Power, Voting, and Voting
  Power: 30 Years After}, pages 1--24. Springer, 2013.

\bibitem{holler2001constrained}
M.J. Holler, R.~Ono, and F.~Steffen.
\newblock Constrained monotonicity and the measurement of power.
\newblock {\em Theory and Decision}, 50(4):383--395, 2001.

\bibitem{holler1983}
M.J. Holler and E.W. Packel.
\newblock Power, luck and the right index.
\newblock {\em Zeitschrift f\"ur National\"okonomie}, 43(1):21--29, 1983.

\bibitem{0083.14301}
J.R. Isbell.
\newblock A class of simple games.
\newblock {\em Duke Math. J.}, 25:423--439, 1958.

\bibitem{johnston1978measurement}
R.J. Johnston.
\newblock On the measurement of power: {S}ome reactions to {L}aver.
\newblock {\em Environment and Planning A}, 10(8):907--914, 1978.

\bibitem{average_representation}
S.~Kaniovski and S.~Kurz.
\newblock The average representation -- a cornucopia of power indices?
\newblock {\em submitted}, page 10 pp., 2014.
\newblock available at http://arxiv.org/abs/1405.0825.

\bibitem{cutting}
V.M. Kartak, S.~Kurz, A.V. Ripatti, and G.~Scheithauer.
\newblock Minimal proper non-irup instances of the one-dimensional cutting
  stock problem.
\newblock {\em Discrete Appl. Math.}, 2014.
\newblock submitted.

\bibitem{krohn1995}
I.~Krohn and P.~Sudh\"olter.
\newblock Directed and weighted majority games.
\newblock {\em Math. Methods Oper. Res.}, 42(2):189--216, 1995.

\bibitem{kurz2012minimum}
S.~Kurz.
\newblock On minimum sum representations for weighted voting games.
\newblock {\em Ann. Oper. Res.}, 196(1):361--369, 2012.

\bibitem{kurz2012inverse}
S.~Kurz.
\newblock On the inverse power index problem.
\newblock {\em Optimization}, 61(8):989--1011, 2012.

\bibitem{inverse}
S.~Kurz.
\newblock The inverse problem for power distributions in committees.
\newblock {\em submitted}, page 46 pp., 2014.
\newblock available at http://arxiv.org/abs/1402.0988.

\bibitem{kurz2012heuristic}
S.~Kurz and S.~Napel.
\newblock Heuristic and exact solutions to the inverse power index problem for
  small voting bodies.
\newblock {\em Ann. Oper. Res.}, 215(1):137--163, 2014.

\bibitem{kurz2013dedekind}
S.~Kurz and N.~Tautenhahn.
\newblock On {D}edekind's problem for complete simple games.
\newblock {\em Internat. J. Game Theory}, 42(2):411--437, 2013.

\bibitem{0243.94014}
S.~Muroga.
\newblock {\em Threshold logic and its applications}.
\newblock New York etc.: Wiley-Interscience, a Division of John Wiley \& Sons,
  Inc. XIV, 478 p., 1971.

\bibitem{riker1962theory}
W.H. Riker.
\newblock {\em The theory of political coalitions}, volume 578.
\newblock Yale University Press New Haven, 1962.

\bibitem{mika1994EC}
M.~Widgr\'en.
\newblock Voting power in the {E}{C} decision making and the consequences of
  two different enlargements.
\newblock {\em European J. Oper. Res.}, 38(5):1153--1170, 1994.

\end{thebibliography}

\end{document}